%% file: arXiv.tex
\documentclass{article}

\usepackage{amsmath}
\usepackage{amssymb}
\usepackage{latexsym}
\usepackage{subfigure}
\usepackage{ifthen}
\usepackage{epic}
\usepackage{calc}
\usepackage{verbatim}
\usepackage{enumerate}
\usepackage{xspace}
\usepackage{array}
\usepackage{graphicx}
\usepackage{a4wide,enumerate,url,multirow,chicago,versions}

\newcommand{\set}[1]{\ensuremath{\left\{ #1 \right\}}}

\newcommand{\card}[1]{\ensuremath{\left| #1 \right|}}

\newcommand{\floor}[1]{\ensuremath{\left\lfloor #1 \right\rfloor}}
\newcommand{\ceil}[1]{\ensuremath{\left\lceil #1 \right\rceil}}

\newcommand{\bigO}[1]{\ensuremath{\mathcal{O}\left(#1\right)}}

\newcommand{\etal}{\emph{et al.}\xspace}

\newcounter{xoffset}
\newcounter{yoffset}

\newcommand{\node}[4]{%
    \put(#2,#3){\circle*{5}}%
    \setcounter{xoffset}{#2}%
    \setcounter{yoffset}{#3}%
    \ifthenelse{\equal{#1}{b}}{\setcounter{yoffset}{#3+5}}{}%
    \ifthenelse{\equal{#1}{l}}{\setcounter{xoffset}{#2+5}}{}%
    \ifthenelse{\equal{#1}{t}}{\setcounter{yoffset}{#3-5}}{}%
    \ifthenelse{\equal{#1}{r}}{\setcounter{xoffset}{#2-5}}{}%
    \put(\thexoffset,\theyoffset){\makebox(0,0)[#1]{\small{#4}}}%
}

\newtheorem{Thm}{Theorem}

\newtheorem{Pro}[Thm]{Proposition}
\newtheorem{Ex}[Thm]{Example}

\newtheorem{Cor}[Thm]{Corollary}

\newtheorem{Rem}[Thm]{Remark}

\roman{mpfootnote}

\includeversion{arxiv}
\excludeversion{tissec}

\newtheorem{Sch}{Construction}

\renewcommand{\etal}{et al.\xspace}

\newcommand{\trlabel}[2]{T_{#1}^{(#2)}}
\newcommand{\dialabel}[3]{D_{#1}^{(#2,#3)}}
\newcommand{\reclabel}[3]{T_{#1,#1}^{(#2,#3)}}

\newenvironment{proof}[1][]
{\begin{trivlist} \item \textbf{Proof #1}~~}{\unskip\hspace*{\fill}\mbox{$\blacksquare$}\par\end{trivlist}}

\begin{document}

\title{Practical and Efficient Cryptographic Enforcement of Interval-Based Access Control Policies}
\author{Jason Crampton\\Royal Holloway, University of London}
\maketitle

\input{abstract}

\input{motivation}

\input{kas_background}

\input{multiplicative3}
\input{geo_spatial3} 
\input{interval1} 

\input{conclusion}

\bibliographystyle{acmtrans}
\bibliography{../../../../workinprogress/index}

\appendix

\input{appendix}

\end{document}

%% file: abstract.tex


\begin{abstract}
The enforcement of access control policies using cryptography has received considerable attention in recent years and the security of such enforcement schemes is increasingly well understood.
Recent work in the area has considered the efficient enforcement of temporal and geo-spatial access control policies, and asymptotic results for the time and space complexity of efficient enforcement schemes have been obtained.
However, for practical purposes, it is useful to have explicit bounds for the complexity of enforcement schemes.

In this paper, we consider interval-based access control policies, of which temporal and geo-spatial access control policies are special cases.
We define enforcement schemes for interval-based access control policies for which it is possible, in almost all cases, to obtain exact values for the schemes' complexity, thereby subsuming a substantial body of work in the literature.
Moreover, our enforcement schemes are more practical than existing schemes, in the sense that they operate in the same way as standard cryptographic enforcement schemes, unlike other efficient schemes in the literature.
The main difference between our approach and earlier work is that we develop techniques that are specific to the cryptographic enforcement of interval-based access control policies, rather than applying generic techniques that give rise to complex constructions and asymptotic bounds.
\end{abstract}

%% file: motivation.tex
\section{Introduction}

In some situations, we may wish to use cryptographic techniques to enforce some form of access control.
Such an approach is useful when data objects have the following characteristics: read often, by many users; written once, or rarely, by the owner of the data; and transmitted over unprotected networks.
\citeN{fu:key06} identify content distribution networks, such as Akami and BitTorrent, as applications where some  kind of cryptographic access control is particularly suitable.
In such circumstances, protected data (objects) are encrypted and authorized users are given the appropriate cryptographic keys.
When cryptographic enforcement is used, the problem we must address is the efficient and accurate distribution of encryption keys to authorized users.

In recent years, there has been a considerable amount of interest in \emph{key encrypting} or \emph{key assignment} schemes.
In such schemes, a user is given a secret value -- typically a single key -- which enables the user to derive some collection of encryption keys which decrypt the objects for which she is authorized.
Key derivation is performed using the secret value and some information made publicly available by the scheme administrator.
%
The two objectives when designing such a scheme are to minimize the amount of public information and the time required to derive a key.
Unsurprisingly, it is not possible to realize both objectives simultaneously, so trade-offs have been sought.
\citeN{cram:csfw06} provide a survey of, and taxonomy for, key assignment schemes, and the various factors that affect the parameters described above.

At the same time, we have seen the development of access control models in which time plays an important role in deciding whether access requests are authorized or not~\cite{bert:trbac01}.
One particular application of such ``temporal access control'' systems is the protection of data that is made available periodically as (part of) a subscription-based service~\cite{bert:temp02}.
Prior to 2006, a number of schemes for enforcing temporal access control policies using cryptographic mechanisms appeared in the literature, many of which have been shown to be insecure (see~\cite{atal:inco07} for a summary of this work).

\citeN{atal:inco07}, \citeN{aten:prov06a} and \citeN{desa:new07} described the first key assignment schemes for temporal access control with provable security properties.
This work focused on two particular aspects:
  \begin{itemize}
    \item the development of schemes that provided key indistinguishability, and
    \item the reduction of the storage required for public information and the number of operations required for key derivation.
  \end{itemize}
One shortcoming of their work is that the methods used to tackle the second of these issues do not consider the actual requirements of the underlying access control policy.
Instead, generic techniques to reduce the diameter of a directed graph are applied.
This has two consequences: optimizations that are tailored to the particular characteristics of the problem are not considered and only the asymptotic behavior of the constructions is provided.
Given that the number of time intervals $m$ is likely to be rather small in many practical applications, it is not clear that this kind of approach is the most appropriate.
Moreover, the absence of explicit bounds means that for small $m$ it is not at all obvious which scheme is optimal.
In short, existing schemes may be efficient (for large values of $m$) but it is questionable whether they are practical.

\citeN{atal:effi07} have also studied the enforcement of ``geo-spatial'' access control policies.
In this context, users are authorized to access data that belongs to particular locations in a rectangular grid.
Atallah \etal\ apply rather similar techniques (as those used for temporal access control policies) to construct asymptotic bounds on the amount of space and the number of derivation steps required.

In this paper, we consider optimizations for both temporal and geo-spatial access control policies that arise from a rather straightforward observation about the particular problem at hand.
This enables us to present concrete schemes with precise bounds on the amount of storage and the number of derivation steps required.

The space and time complexity  of cryptographic enforcement schemes can be measured in ``edges'' and ``hops'' respectively~\cite{cram:csfw06}.
For the enforcement of a temporal access control policy with $m$ time points, for example, we require $m(m-1)$ edges and $\ceil{\log_2 m}$ hops\footnote{Henceforth, all logarithms are base $2$, unless explicitly stated otherwise.}, whereas \citeN{atal:inco07} require $\bigO{m^2}$ edges and $\bigO{\log^* m}$ hops and \citeN{desa:new08} require $\bigO{m^2 \log m}$ edges and $\bigO{\log^* m}$ hops.%
\footnote{The function $\log^* : \mathbb{N} \rightarrow \mathbb{N}$ is the \emph{iterated log function}, where $\log^* m = 0$ if $m \leqslant 1$ and \mbox{$\log^* m = 1 + \log^* (\log m)$} for $m > 1$. The iterated log function grows very slowly: $\log^* m \leqslant 4$ for all $m \leqslant 2^{16}$ and $\log^* m \leqslant 5$ for all $m \leqslant 2^{65536}$, for example.}
However, substantial multiplicative constants and lower-order terms may be hidden by the $\mathcal{O}$ notation when it comes to the number of edges and the number of hops required for key derivation.%
\footnote{The schemes in the literature do not consider the multiplicative constants or lower-order terms.  It is, perhaps, an indication of the complexity of the constructions in the literature that we have not, despite considerable effort, been able to determine the multiplicative constants in the expressions given for the number of edges.}

For values of $m$ that are likely to be used in practice, these terms will be of considerable importance.
The actual number of hops required by the scheme of \citeN{atal:inco07}, for example, is $2\log^* m + 4$, which, for many values of $m$ of practical interest, will be greater than $\ceil{\log m}$.
Moreover, the edge sets that are used in existing efficient constructions require bespoke algorithms for key derivation and modifications to the basic operation of a key assignment scheme~\cite{atal:effi07}.
In contrast, key derivation for our constructions remains very simple.

Finally, we demonstrate that temporal and geo-spatial access control policies (at least as they are understood in the context of key assignment schemes) are special cases of a more general type of policy, which we call an \emph{interval-based access control policy}.
Such policies are parameterized by an integer $k$, where temporal and geo-spatial policies correspond to the cases $k=1$ and $k=2$, respectively.
Perhaps the most important contribution of this paper is to describe how to construct a set of edges for an arbitrary value of $k$ and provide tight bounds on the number of edges and key derivation hops required.

In summary, the main contributions of this paper are:
  \begin{itemize}
    \item to generalize the problem of enforcement of temporal and geo-spatial access control policies to the enforcement of interval-based access control policies;
    \item to provide tight bounds on the complexity of enforcing temporal, geo-spatial and interval-based access control policies using key assignment schemes;
    \item to provide simple, concrete constructions for such schemes.
  \end{itemize}

In the next section, we describe some relevant background material, define what we mean by an interval-based access control policy, and introduce the problem of enforcing an interval-based access control policy using cryptographic mechanisms.
In Section~\ref{sec:temporal}, we consider temporal access control policies.
The main contribution of this section is to state and prove a rather general result and explore some special cases of this result.
In this section we also consider constructions in which the user may have more than one key.
In Section~\ref{sec:geo-spatial}, we consider the related problem of cryptographic enforcement of geo-spatial access control policies.
We describe relevant related work in both Section~\ref{sec:temporal} and~\ref{sec:geo-spatial}.
In Section~\ref{sec:interval}, we derive results for general interval-based access control policies.
We conclude the paper with a summary of our contributions and some suggestions for future work.

%% file: kas_background.tex
\section{Key Assignment Schemes}\label{sec:background}

Given a partially ordered set of security labels $(L,\leqslant)$, an \emph{information flow policy} requires that each user $u$ and protected object $o$ be assigned a security label and that information flows between objects and users are consistent with the ordering $\leqslant$: specifically, $u$ is authorized to read $o$ provided the security label of $u$ is greater than or equal to that of $o$~\cite{bell:secu76}.
More formally, let $\lambda : U \cup O \rightarrow L$ be a labeling function that associates each entity with a security label.
Then $u$ is authorized to access $o$ if and only if $\lambda(u) \geqslant \lambda(o)$.

A \emph{key assignment scheme} may be used to enforce an information flow policy.
In such a scheme, it is assumed that every node in $L$ is associated with a symmetric cryptographic key.
For a given node $x \in L$, all objects associated with $x$ are encrypted with the appropriate key, and all users associated with $x$ are given, or can derive, the key for node $x$ and for each node less than or equal to $x$.

More formally, a key assignment scheme comprises a set of keys $\set{\kappa(x) : x \in L}$ and a set of public information.
Each object with security label $x$ is encrypted with $\kappa(x)$.
A user $u$ with the key $\kappa(\lambda(u))$ must be able to derive $\kappa(y)$ for any $y \leqslant \lambda(u)$, using $\kappa(\lambda(u))$ and public information.
Hence, a user can decrypt any object with security label $\lambda(y)$, where $y \leqslant \lambda(u)$.
The first such scheme was described by~\citeN{akl:cryp83}.
The parameters that characterize the behavior of a key assignment scheme are:
  \begin{itemize}
    \item the number of keys that a user requires;
    \item the amount of public information that is required;\footnote{The public information always includes a data structure encoding $(L,\leqslant)$.}
    \item the amount of time taken to derive a key (equivalently, the number of operations required to perform key derivation).
  \end{itemize}

We could, trivially, give a user the key associated with each label for which she is authorized, but this type of approach is rarely considered appropriate.
Most of the literature on key assignment schemes assumes that each user has a single secret value and the keys for which she is authorized are derived from this secret value.
In general, the more public information employed by the scheme, the smaller the number of key derivation steps required in the worst case.

\subsection{Correctness and security}

A key assignment scheme that enforces an information flow policy for $(L,\leqslant)$ must be correct and it must be secure.
Informally, we say a key assignment scheme is
  \begin{itemize}
    \item correct if each user can derive the keys for which she is authorized;
    \item secure if no set of users can derive a key for which none of them is authorized.
  \end{itemize}
Recently, the notions of \emph{key recovery} and \emph{key indistinguishability} have been introduced to capture in more formal terms what it means for a key assignment scheme to be secure~\cite{atal:dyna09,aten:prov06a}.
Informally, to obtain a scheme with the key recovery property, each node $x \in V$ is associated with a secret value $\kappa(x)$, and, for each edge $(x,y) \in E$, we publish $Enc_{\kappa(x)}(\kappa(y))$, where $Enc_{\kappa}(M)$ denotes the encryption of message $M$ using key $\kappa$.
%
Then any user in possession of $\kappa(x)$ can derive $\kappa(y)$ in one step, and for any $z$ on a path from $x$ containing $e$ edges, $\kappa(z)$ can be (iteratively) derived in $e$ steps.
Such a scheme can be extended to one with the property of key indistinguishability by associating a secret value $\sigma(x)$ with each node $x$, making $\kappa(x)$ a function of $\sigma(x)$ and using $\sigma(x)$ to derive $\sigma(y)$.

For the purposes of this paper, it is sufficient to note that given a directed, acyclic graph $G = (V,E)$, there exists a key assignment scheme that has the property of key indistinguishability, the amount of storage required is proportional to $\card{E}$ (the cardinality of $E$), and the number of derivation steps required is equal to the diameter of $G$ (the length of the longest path in $G$).
The interested reader is referred to the literature for further details~\cite{atal:dyna09,aten:prov06a}.

\subsection{Derivation-storage trade-offs}

A partially ordered set $(L,\leqslant)$ can be represented by a directed, acyclic graph $(V,E)$, where $V = L$.
There are two obvious choices for the edge set $E$: one is the full partial order relation $\leqslant$; the second is to omit all transitive and reflexive edges from $\leqslant$ to obtain the \emph{covering relation}, denoted $\lessdot$.
The graph $(L,\lessdot)$ is called the \emph{Hasse diagram} of $L$, and is the standard representation of $L$ as a directed graph~\cite{dave:intr02}.

It can be seen that a key assignment scheme for a directed graph can be used specifically to enforce an information flow policy.
We may use the graph $(L,\leqslant)$, in which case key derivation can always be performed in one step.
In contrast, key derivation may require a number of steps when we use the graph $(L,\lessdot)$.
The trade-off here is that the second graph contains fewer edges and hence the number of items of public information that are required to support key derivation is smaller.
The study of these kinds of trade-offs will be the focus of this paper.

\subsection{Interval-based access control policies}

Let $O$ be a set of protected objects, let $U$ be a set of users, and let $A_1,\dots,A_k$ be finite, totally ordered sets of cardinality $n_1,\dots,n_k$, respectively.
We write $\mathcal{A}$ to denote $\prod_{i=1}^k A_i = A_1 \times \dots \times A_k$.

We say $[x_i,y_i] \subseteq A_i$, where $1 \leqslant x_i \leqslant y_i \leqslant n_i$, is an \emph{interval} in $A_i$.
We say $\prod_{i=1}^k [x_i,y_i] = [x_1,y_1] \times \dots \times [x_k,y_k] \subseteq \mathcal{A}$ is a \emph{hyperrectangle}.
We write $\mathsf{HRec}(\mathcal{A})$ to denote the set of hyperrectangles in $\mathcal{A}$.

We assume that each object $o \in O$ is associated with a unique attribute tuple $(a_1,\dots,a_k) \in \mathcal{A}$, and each user $u \in U$ is authorized for some hyperrectangle $\prod_{i=1}^k [x_i,y_i] \in \mathsf{HRec}(\mathcal{A})$.
Then we say that a user $u$ associated with $\prod_{i=1}^k [x_i,y_i]$ is \emph{authorized} to read an object $o$ associated with tuple $(a_1,\dots,a_k) \in \mathcal{A}$ if and only if $a_i \in [x_i,y_i]$ for all $i$.
Such a policy may be enforced using cryptographic methods:
  \begin{itemize}
    \item each attribute tuple $a = (a_1,\dots,a_k) \in \mathcal{A}$ is associated with a cryptographic key, which we denote by $\kappa(a)$;
    \item all objects $o$ that are associated with $a$ are encrypted with $\kappa(a)$;
    \item $u$ should be able to derive $\kappa(a)$ whenever $a_i \in [x_i,y_i]$ for all $i$.
  \end{itemize}

The problem that we consider in the remainder of this paper is the construction of a set of edges $E$ for the set of nodes $\mathsf{HRec}(\mathcal{A})$ such that:
  \begin{itemize}
    \item for all $\prod_{i=1}^k [x_i,y_i]$ and all $(a_1,\dots,a_k)$, there exists a path from $\prod_{i=1}^k [x_i,y_i]$  to $(a_1,\dots,a_k)$ if and only if $a_i \in [x_i,y_i]$.
    \item $\card{E}$ is small;
    \item the diameter of the graph $(\mathsf{HRec}(\mathcal{A}),E)$ is small.
  \end{itemize}
The first criterion requires that the graph implements the desired access control policy.
We say a set of edges $E$ is \emph{policy-enforcing}, or simply \emph{enforcing}, if it satisfies this criterion.
The second means that we wish to keep the public storage requirements small, while the final criterion requires that the complexity of worst-case key derivation time be low.

In the remainder of this section we review two special cases of interval-based access control that have been widely studied in the literature.
To simplify our exposition and comparison with related work, we will consider these special cases in detail in Sections~\ref{sec:temporal} and~\ref{sec:geo-spatial}, before studying the general case in Section~\ref{sec:interval}.

\subsubsection{Temporal access control}

When $k=1$, we have $\mathcal{A} = A_1$.
It is customary to interpret $A_1$ as a finite set of $n$ consecutive time points (see~\cite{atal:inco07,desa:new08}, for example).
Each object is associated with a unique time point, and each user is associated with a set of consecutive time points (an \emph{interval}).
Without loss of generality, we assume that the time points are in one-to-one correspondence with the integers $1,\dots,n$.
We write $[x,y]$ to denote the set $\set{t : x \leqslant t \leqslant y}$.
Then each object is associated with some integer $x \in [1,n]$ and each user is associated with some interval $[x,y] \subseteq [1,n]$.
A user associated with interval $[x,y]$ should be able to derive $\kappa(t)$ for all $t \in [x,y]$.

\begin{figure}[h] \centering
  \includegraphics{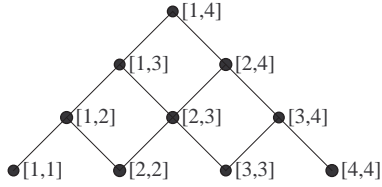}
\caption{The Hasse diagram of $(T_4,\subseteq)$} \label{fig:intervals-as-hasse-diagram}
\end{figure}

Henceforth, we write $T_n$ to denote the set of intervals in $[1,n]$: that is,
  \[
    T_n \stackrel{\rm def}{=} \set{[x,y] : 1 \leqslant x \leqslant y \leqslant n}.
  \]
We denote the set of all intervals by $T_n$ because the partially ordered set $(T_n,\subseteq)$ has a natural representation as a triangular grid, as illustrated in Figure~\ref{fig:intervals-as-hasse-diagram}.
We may refer to $T_n$ as an $n$-triangle.
A node of the form $[x,x] \in T_n$ is equivalent to a point $x \in [1,n]$ and will be called a \emph{leaf node}.
The set of leaf nodes corresponds to the totally ordered set of time points $1,\dots,n$.

\subsubsection{Geo-spatial access control}

When $k = 2$, we have $\mathcal{A} = A_1 \times A_2$, which represents a finite rectangular grid of points.
In this case each object is associated with a unique point in the grid, and each user is associated with a set of points that correspond to a sub-rectangle of the rectangular grid~\cite{atal:effi07}.
Without loss of generality, we assume $A_1 = \set{1,\dots,m}$ and $A_2 = \set{1,\dots,n}$.
Then each object is associated with some point $(x,y)$ and each user is associated with some rectangle $[x_1,y_1] \times [x_2,y_2] = \set{(t_1,t_2) : t_1 \in [x_1,y_1], t_2 \in [x_2,y_2]}$.

We write $T_{m,n}$ (as an abbreviation of the more accurate $T_m \times T_n$) to denote the set of rectangles defined by a rectangular $m \times n$ grid of points: that is
\[
  T_{m,n} \stackrel{\rm def}{=} \set{[x_1,y_1] \times [x_2,y_2] : 1 \leqslant x_1 \leqslant y_1 \leqslant m, 1 \leqslant x_2 \leqslant y_2 \leqslant n}.
\]
Nodes of the form $[x,x] \times [y,y]$~--~which may also be interpreted as the point $(x,y)$~--~will be called \emph{leaf nodes}.
The set of leaf nodes corresponds to the set of points in the rectangular $m \times n$ grid.

\begin{figure}[h]
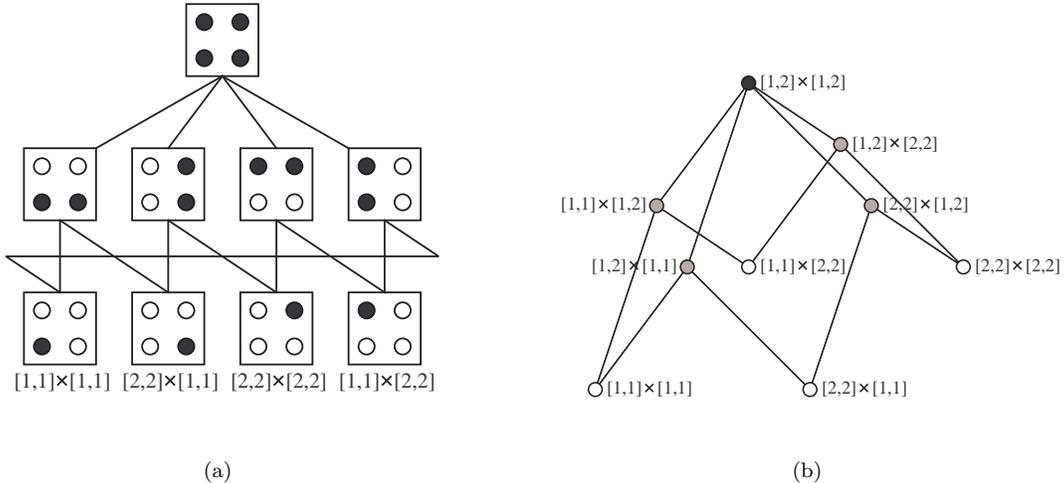
 \centering
  \subfigure[]{\begin{minipage}[b]{.45\textwidth}\centering\includegraphics[width=\textwidth]{t_2_2_poset.md}\end{minipage}}
  \hfill
  \subfigure[]{\begin{minipage}[b]{.5\textwidth}\centering\includegraphics[width=\textwidth]{t_2_2.md}\end{minipage}}
\caption{Two representations of $T_2 \times T_2$} \label{fig:t-2-times-t-2}
\end{figure}

It is rather difficult to represent $T_{m,n}$ in two dimensions for all but the smallest values of $m$ and $n$.
Two different visualizations of $T_{2,2}$ are shown in Figure~\ref{fig:t-2-times-t-2}: the first simply illustrates it as a partially ordered set of subsets ordered by subset inclusion in which rectangles are represented by filled circles; the second illustrates it by building the rectangles on top of a $2 \times 2$ grid (in a manner analogous to the representation of $T_m$ used in Figure~\ref{fig:intervals-as-hasse-diagram}).
In the second figure, nodes of the same color have the same area (as rectangles): all rectangles of area $2$ are filled in gray, whereas all rectangles of area $1$ are filled white.
Although the first visualization is perhaps easier to interpret, it is the second visualization that we will have in mind when developing our constructions in Section~\ref{sec:geo-spatial}.

%% file: multiplicative3.tex
\section{Temporal Access Control}\label{sec:temporal}

In this section we first describe two rather simple schemes that will be used as ``building blocks'' for more complex schemes.
Then, in Section~\ref{sec:temporal-single-key}, we describe a general construction in which users have a single key, and derive a number of concrete constructions as special cases.
In particular, we describe a construction for $T_m$ in which the resulting graph has $\bigO{m^2 \log \log m}$ edges and diameter $\log \log m$.
In Section~\ref{sec:temporal-multi-key} we describe a construction in which users may have two keys.

Before proceeding any further, we note the existence of a lower bound on the cardinality of an enforcing set of edges and the existence of an enforcing set of edges that yields a graph of diameter~$1$.

\begin{Pro}\label{pro:minimum-cardinality-enforcing-set}
Let $E$ be an enforcing set of edges for $T_m$.  Then \mbox{$\card{E} \geqslant m(m-1)$}.
\end{Pro}

\begin{proof}
Suppose that $E$ is an enforcing set of edges such that $\card{E} < m(m-1)$.
Then at least one non-leaf node $[x,y]$, where $y > x$, has out-degree less than $2$.
This implies one of two things:
  \begin{itemize}
    \item either there exists $z \in [x,y]$ such that $[z,z]$ is not reachable from $[x,y]$;
    \item or there exists an edge from $[x,y]$ to $[x',y']$ such that $[x,y] \ne [x',y']$ and all $z \in [x,y]$ are reachable from $[x',y']$.
  \end{itemize}
In the first case, the edge set does not satisfy the requirement that $[z,z]$ is reachable from $[x,y]$ if $z \in [x,y]$.
In the second case, there are two possibilities:
  \begin{itemize}
    \item either $[x,y] \subset [x',y']$, in which case there exists $z \in [x',y']$ such that $z \not\in [x,y]$ and $z$ is reachable from $[x,y]$, contradicting the requirement that $z$ is reachable from $[x,y]$ only if $z \in [x,y]$; or
    \item $[x,y] \not\subset [x',y']$, so there exists $z \in [x,y]$ such that $z \not\in [x',y']$ and $[z,z]$ is reachable from $[x',y']$, which contradicts the requirement that $z$ should be reachable only if $z \in [x',y']$.
  \end{itemize}
The result follows.
\end{proof}

\begin{tissec}
\begin{Pro}[(\citeN{cram:nordsec09})]\label{pro:1-hop-scheme}
There exists an enforcing set of edges $E$ such that $\card{E} = \frac{1}{6}m(m-1)(m+4)$ and the diameter of $(T_m,E)$ is $1$.
\end{Pro}
\end{tissec}

\begin{arxiv}
\begin{Pro}[\citeN{cram:nordsec09}]\label{pro:1-hop-scheme}
There exists an enforcing set of edges $E$ such that $\card{E} = \frac{1}{6}m(m-1)(m+4)$ and the diameter of $(T_m,E)$ is $1$.
\end{Pro}
\end{arxiv}

\begin{tissec}
\begin{Pro}[(\citeN{cram:nordsec09})]\label{pro:log-m-hop-scheme}
There exists an enforcing set of edges $E$ such that $\card{E} = m(m-1)$ and the diameter of $(T_m,E)$ is $\ceil{\log m}$.
\end{Pro}
\end{tissec}

\begin{arxiv}
\begin{Pro}[\citeN{cram:nordsec09}]\label{pro:log-m-hop-scheme}
There exists an enforcing set of edges $E$ such that $\card{E} = m(m-1)$ and the diameter of $(T_m,E)$ is $\ceil{\log m}$.
\end{Pro}
\end{arxiv}

To establish the above result, we now describe a construction, which we call \emph{binary decomposition}, that generates a set of edges with the stated properties.
Binary decomposition is a generalization of a construction we presented in an earlier paper~\cite[Scheme 2]{cram:nordsec09}.
Binary decomposition is optimal, in the sense that any set of enforcing edges must have cardinality at least $m(m-1)$ (by Proposition~\ref{pro:minimum-cardinality-enforcing-set}).
We first introduce some additional notation: we write $D_n$ to denote an $n$-diamond, which is formed by joining two copies of $T_n$ along the long diagonal; and we write $R_{m,n}$ to denote a rectangular grid of nodes of side lengths $m$ and $n$.

\begin{figure*}[!ht]
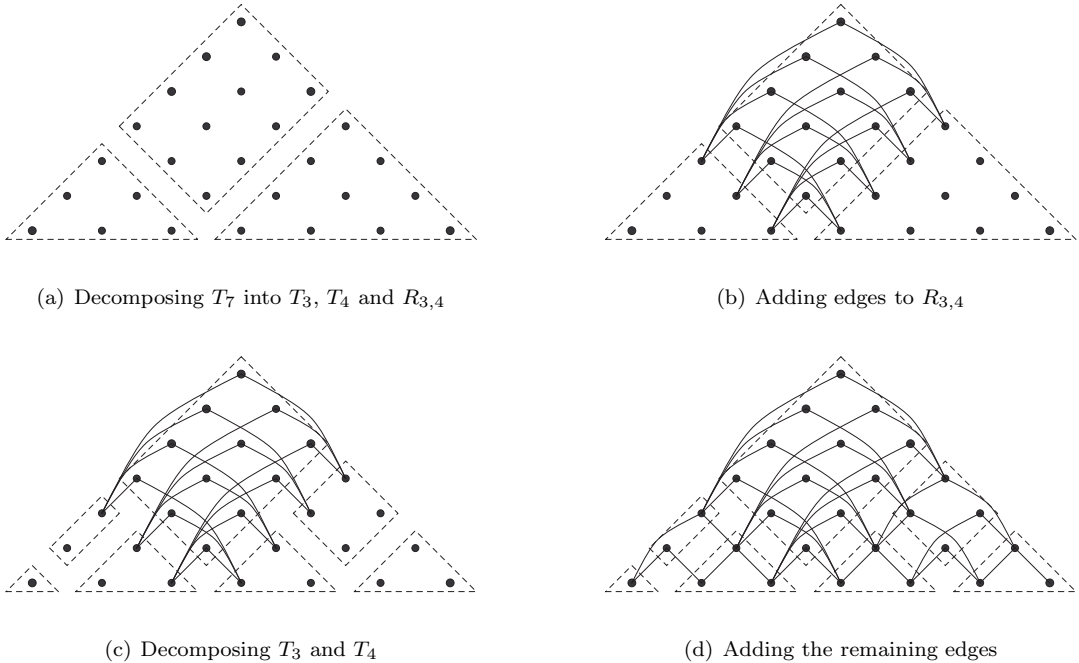

  \subfigure[Decomposing $T_{7}$ into $T_3$, $T_4$ and $R_{3,4}$]{ \includegraphics[width=.45\textwidth]{t_7_decomposition.md} } \hfill
  \subfigure[Adding edges to $R_{3,4}$]{ \includegraphics[width=.45\textwidth]{t_7_first_step.md} }
  \subfigure[Decomposing $T_3$ and $T_4$]{ \includegraphics[width=.45\textwidth]{t_7_middle_step.md} } \hfill
  \subfigure[Adding the remaining edges]{  \includegraphics[width=.45\textwidth]{t_7_last_step.md} }
\caption{The binary decomposition of $T_7$} \label{fig:t-11-construction}
\end{figure*}

\begin{tissec}
\begin{Sch}[(Binary decomposition)]\label{sch:log-m-derivation}
Let $\ell = \floor{m/2}$ and $r = \ceil{m/2}$.
Now $T_m$ comprises:
  \begin{itemize}
    \item a copy of $T_{\ell}$, containing the minimal elements $[1,1],\dots,[\ell,\ell]$;
    \item a copy of $T_{r}$, containing the minimal elements $[\ell + 1,\ell + 1],\dots,[m,m]$;
    \item a copy of rectangle $R_{\ell,r}$, containing the remaining nodes in $T_m$.
  \end{itemize}
This view of $T_7$ is depicted in Figure~\ref{fig:t-11-construction}(a).
Notice that every interval represented by a node in $R_{\ell,r}$ contains $\ell$ and $\ell + 1$.

The first step in the construction of $E$, then, is to include an edge from every node in $R_{\ell,r}$ to one node in $T_{\ell}$ and one node in $T_{r}$.
In particular, for node $[x,y]$ such that $x \leqslant \ell < y$, we add edges from $[x,y]$ to $[x,\ell]$ and from $[x,y]$ to $[\ell+1,y]$.

We now recursively apply this construction to $T_{\ell}$ and $T_{r}$, terminating when $\ell,r \leqslant~1$.
The construction of the edge set for $T_{7}$ is illustrated in Figure~\ref{fig:t-11-construction}.
\end{Sch}
\end{tissec}

\begin{arxiv}
\begin{Sch}[Binary decomposition]\label{sch:log-m-derivation}
Let $\ell = \floor{m/2}$ and $r = \ceil{m/2}$.
Now $T_m$ comprises:
  \begin{itemize}
    \item a copy of $T_{\ell}$, containing the minimal elements $[1,1],\dots,[\ell,\ell]$;
    \item a copy of $T_{r}$, containing the minimal elements $[\ell + 1,\ell + 1],\dots,[m,m]$;
    \item a copy of rectangle $R_{\ell,r}$, containing the remaining nodes in $T_m$.
  \end{itemize}
This view of $T_7$ is depicted in Figure~\ref{fig:t-11-construction}(a).
Notice that every interval represented by a node in $R_{\ell,r}$ contains $\ell$ and $\ell + 1$.

The first step in the construction of $E$, then, is to include an edge from every node in $R_{\ell,r}$ to one node in $T_{\ell}$ and one node in $T_{r}$.
In particular, for node $[x,y]$ such that $x \leqslant \ell < y$, we add edges from $[x,y]$ to $[x,\ell]$ and from $[x,y]$ to $[\ell+1,y]$.

We now recursively apply this construction to $T_{\ell}$ and $T_{r}$, terminating when $\ell,r \leqslant~1$.
The construction of the edge set for $T_{7}$ is illustrated in Figure~\ref{fig:t-11-construction}.
\end{Sch}
\end{arxiv}

\subsection{Single-key constructions} \label{sec:temporal-single-key}

\citeN[\S 5.2]{cram:nordsec09} described a construction that generates an enforcing set of edges $E$ of cardinality $\frac{1}{6}m(m-1)(\sqrt{m}+4)$ such that the diameter of $(T_m,E)$ is~$2$.
The construction was based on a factorization of $m$ into two integers.
We now prove a more general result, in which we express $m$ as the product of $d$ integers and construct a set of enforcing edges $E$ such that the diameter of $(T_m,E)$ is $d$.
The result has a number of interesting corollaries, which we also explore in this section.

\begin{Thm} \label{thm:multiplicative-decomposition-repeated-factors}
Let $m = \prod_{i=1}^d a_i$, where $a_i$ is an integer and $2 \leqslant a_i \leqslant a_{i+1}$ for all $i$.
Then there exists an enforcing set of edges $E$ such that
  \begin{equation}\label{eq:enforcing-edges-mult-decomposition}
    \card{E} = \frac{m^2}{6} \sum_{i=1}^d \frac{(a_i-1)(a_i+4)}{\pi_i},
  \end{equation}
where $\pi_i = \prod_{j=1}^i a_j$, and the diameter of $(T_m,E)$ is $d$.
\end{Thm}

The result is proved by induction on $d$ and by partitioning $T_m$ into ``supernodes'', which are copies of smaller triangles and diamonds.
Informally, the inductive step works by splitting $T_m$ into a triangle $T_a$ of supernodes, where $m = ab$, each non-leaf supernode is a copy of $D_b$ and each leaf supernode is a copy of $T_b$.
The application of the construction to $T_{12}$, where $a = 3$ and $b = 4$, is illustrated in Figure~\ref{fig:splitting}.

\begin{proof}[of Theorem~\ref{thm:multiplicative-decomposition-repeated-factors}]
First consider the case $d = 1$.
By Proposition~\ref{pro:1-hop-scheme}, there exists an edge set with cardinality $\frac{1}{6}m(m-1)(m+4)$ and the diameter of the graph is $1 = d$.
Note that for $d = 1$, we have $a_1 = m$ and $\pi_1 = m$ in~\eqref{eq:enforcing-edges-mult-decomposition}, so the result holds for $d=1$.%

Now let us assume the result holds for all $d < D$ and consider $m = \prod_{i=1}^D a_i$.
For convenience, we write $b = \prod_{i=2}^D a_i$ (that is, $m = a_1 b$).
We first note that every node in $T_m$ can be written in the form $[x+\alpha b,y+\beta b]$, where $1 \leqslant x,y \leqslant b$ and $0 \leqslant \alpha \leqslant \beta < a_1$.
  \begin{itemize}
    \item If $\alpha = \beta$, then $[x+\alpha b,y+\beta b]$ belongs to a leaf triangle supernode which we denote by $\trlabel{b}{\alpha}$, $0 \leqslant \alpha < a_1$.
    \item If $\alpha < \beta$, then $[x+\alpha b,y+\beta b]$ belongs to a non-leaf diamond supernode, which we denote by $\dialabel{b}{\alpha}{\beta}$, $0 \leqslant \alpha < \beta < a_1$.
  \end{itemize}
Note also that $[x+\alpha b,y+\beta b]$, where $\alpha < \beta$, is the (disjoint) union of the intervals
  \[
    [x+\alpha b, b + \alpha b],[1 + (\alpha + 1)b,b + (\alpha+1)b],\dots,[1+(\beta-1)b, b + (\beta - 1) b],[1+\beta b, y + \beta b],
  \]
which belong, respectively, to $\trlabel{b}{\alpha},\trlabel{b}{\alpha+1},\dots,\trlabel{b}{\beta-1},\trlabel{b}{\beta}$.

Now consider a node $[x,y] \in \dialabel{b}{0}{a_1-1}$.
(In other words, $[x,y]$ belongs to the maximal supernode in $T_{a_1}$.)
Then for each such $[x,y]$, we define
  \[
    C_{x,y}  \stackrel{\rm def}{=} \set{[x + \alpha b , y + \beta b] : 0 \leqslant \alpha < \beta < a_1}.
  \]
Then each diamond supernode contains exactly one element of $C_{x,y}$ and the elements of $C_{x,y}$ (ordered by subset inclusion) form a copy of $T_{a_1-1}$.
Since $[x+\alpha b, y + \beta b]$ is the union of intervals in $\trlabel{b}{\alpha},\dots,\trlabel{b}{\beta}$, we can connect the nodes in $C_{x,y}$ directly to the appropriate leaf nodes using a $1$-hop construction for $T_{a_1}$, which requires $\frac{1}{6}a_1(a_1-1)(a_1+4)$ edges.
Moreover, $C_{x,y} = C_{x',y'}$ if and only if $x = x'$ and $y = y'$.
Hence, there are $b^2 = m^2/a_1^2$ different sets $C_{x,y}$ (since there are $b$ choices for each of $x$ and $y$).
%
%
Hence, we create a total of $\frac{m^2}{6a_1}(a_1-1)(a_1+4)$ edges, which enable us to jump directly from any node in a diamond supernode to some node in a leaf supernode.
We denote this set of edges by $E_{\rm outer}$.

By the inductive hypothesis, there exists an enforcing set of edges $E_{\rm inner}$ for $\trlabel{b}{\alpha}$, $0 \leqslant \alpha < a_1$.
Moreover, since $m/a_1 = b = \prod_{i=2}^D a_i$ and $\trlabel{b}{\alpha}$ is a copy of $T_{m/a_1}$, we have by the inductive hypothesis
  \[
    \card{E_{\rm inner}} = \frac{m^2}{6a^2_1}\sum_{i=2}^D \frac{(a_i-1)(a_i+4)}{a_2\dots a_i}.
  \]
In total, we require $\card{E_{\rm outer}} + a_1 \card{E_{\rm inner}}$ edges (since there are $a_1$ copies of $T_b$).
Hence the number of edges is given by
  \[
    \frac{m^2}{6a_1}(a_1-1)(a_1+4) + \frac{m^2}{6a_1}\sum_{i=2}^D \frac{(a_i-1)(a_i+4)}{a_2\dots a_i} = \frac{m^2}{6}\sum_{i=1}^D \frac{(a_i-1)(a_i+4)}{a_1\dots a_i},
  \]
as required.

Moreover, it is clear that the resulting set of edges $E$ is enforcing if $E_{\rm inner}$ is enforcing (which it is, by the inductive hypothesis).
Finally, the diameter of $(\trlabel{b}{\alpha},E_{\rm inner})$ is $D-1$ by the inductive hypothesis.
Hence, the diameter of $(T_m,E)$ is $1 + (D-1) = D$, as required.
\end{proof}

\begin{Ex}
The partial construction of a two-hop scheme for $T_{12}$ is illustrated in Figure~\ref{fig:splitting}.
We divide $T_{12}$ into copies of $D_4$ and $T_4$, yielding a copy of $T_3$ in which the non-leaf supernodes are diamonds and leaf supernodes are triangles (as depicted in Figure~\ref{fig:splitting}(a)).

\begin{figure}[!th]
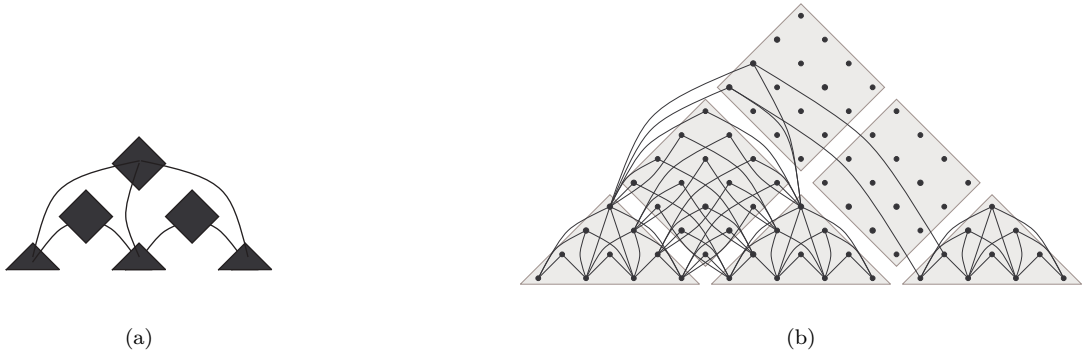
 \centering
  \subfigure[]{
      \includegraphics{t_4_as_diamonds.md}
  }
    \hfill
  \subfigure[]{
    \includegraphics[scale=.45]{t_12_3_4.md}
  }
  \caption{Creating a $2$-hop scheme for $T_{12}$ using $1$-hop schemes for $T_3$ and $T_4$}\label{fig:splitting}
\end{figure}

A one-hop construction for $T_3$ requires seven edges, and must be duplicated for every node in the root supernode (and there are $4^2 = 16$ such nodes).
Hence we require $7 \cdot 16 = 112$ edges to connect nodes in non-leaf supernodes to nodes in leaf supernodes.
(A subset of these edges is depicted in Figure~\ref{fig:splitting}(b).)
Having done this, we can now get from any node that is contained in a copy of $D_4$ to a node in $T_4$ in one hop.

It remains, therefore, to construct an edge set for each $T_4$ supernode such that we can get from any non-leaf node to a leaf node in one hop.
We require $16$ edges for each of the three copies of $T_4$ (a total of $48$ edges).
The construction therefore generates a total of $160$ edges.
\end{Ex}

In the statement of Theorem~\ref{thm:multiplicative-decomposition-repeated-factors}, note that $a_i \geqslant 2$, so
  \[
    m  = \prod_{i=1}^d a_i \geqslant 2^d\qquad \text{and}\qquad d \leqslant \log m.
  \]
Note also that the $i$th term in the summation
  \[
    \frac{(a_i-1)(a_i+4)}{\pi_i} = \frac{1}{\pi_{i-1}}\left(a_i+3-\frac{4}{a_i}\right)
  \]
is minimized when $a_i = 2$.
Finally, note that the difference between successive terms in the summation is given by
  \[
    \frac{(a_{i+1}-1)(a_{i+1}+4)}{a_{i+1}} - (a_i -1)(a_i+4),
  \]
which is approximately equal to zero when $a_{i+1} \approx a_i^2$.
These observations lead to two corollaries of Theorem~\ref{thm:multiplicative-decomposition-repeated-factors}.
The first corollary provides a concise characterization of the number of edges required for a $d$-hop solution when $m = a^d$ for some integer $a$ (which itself includes binary decomposition as a special case).
The second of these results provides an explicit bound for the number of edges required in a scheme with $\log\log m$ steps.

\begin{Cor}\label{cor:a^d}
If $m = a^d$, then there exists an enforcing edge set $E$ such that $\card{E} = \frac{1}{6}m(m-1)(a+4)$ and the diameter of $(T_m,E)$ is $d = \log_a m$.
In particular, if $m = 2^d$, then there exists an edge set of cardinality $m(m-1)$ and a graph of diameter $\ceil{\log m}$.
\end{Cor}

\begin{proof}
By Theorem~\ref{thm:multiplicative-decomposition-repeated-factors}, we have
  \begin{align*}
    \card{E} &= \frac{m^2}{6} (a-1)(a+4) \sum_{i=1}^d \frac{1}{a^i} \\
        &= \frac{m^2}{6} (a-1)(a+4) \left(\frac{1}{a-1}\right)\left(1 - \frac{1}{a^d}\right) \\
        &= \frac{m^2}{6} (a+4) \left(\frac{m-1}{m}\right) \\
        &= \frac{1}{6}m(m-1)(a+4).
  \end{align*}
And for $a = 2$, we have $\card{E} = m(m-1)$.
\end{proof}

\begin{Cor}
Let $m = 2^{2^d}$ for some integer \mbox{$d \geqslant 2$}.
Then there exists an enforcing edge set $E$ such that
  \[
    \card{E} < m^2 \left(1+\frac{1}{6}\log\log m\right)
  \]
and the diameter of $(T_m,E)$ is $\log \log m$.
\end{Cor}

\begin{proof}
We write $m = 2^2 2^{2^1} 2^{2^2} 2^{2^3} \dots 2^{2^{d-1}}$ and apply Theorem~\ref{thm:multiplicative-decomposition-repeated-factors}, thereby obtaining a $d$-hop scheme.
In particular, we have $a_1 = 2^{2^1}$, $a_i = 2^{2^{i-1}}$, $i \geqslant 2$, and $\pi_i = 2^{2^i}$, $i \geqslant 1$.
Hence,
\begin{align*}
	\card{E}
    &= \frac{m^2}{6}\left(\frac{3 \cdot 8}{4} + \sum_{i=2}^d \frac{(2^{2^{i-1}}-1)(2^{2^{i-1}} + 4)}{2^{2^i}}\right) \\
    &= \frac{m^2}{6} \left(6 + \sum_{i=2}^{d} \left(1 + \frac{3}{2^{2^{i-1}}} - \frac{4}{2^{2^i}}\right)\right) \\
    &< \frac{m^2}{6} \left(6 + (d-1) + 3\sum_{i=1}^{d-1} \frac{1}{2^{2^i}} \right) \\
    &< \frac{m^2}{6} \left(6 + (d-1) + 3\sum_{i=1}^{\infty} \frac{1}{4^i} \right) \\ 
    &= \frac{m^2}{6} \left(6 + d\right) \\
    &= m^2 \left(1 + \frac{1}{6}\log \log m \right)
\end{align*}
as required.
\end{proof}

\begin{Rem}
It is worth noting that $\log\log m \leqslant 6$ for all $m \leqslant 2^{2^6} = 2^{64}$.
In other words, for all practical values of $m$, there exists an enforcing set of edges whose cardinality is bounded by $2m^2$ and for which the number of derivation hops is bounded by $\log\log m$.
\end{Rem}

\subsection{Multi-key constructions}\label{sec:temporal-multi-key}

In this section, we consider the trade-off that is possible when we assume that users may have two secret keys (rather than one).
In Appendix~\ref{app:multi-key-temporal}, we consider the additional trade-offs that are possible when the user may have more than two keys.
The basic idea is to define a set of special nodes $T'_m \subseteq T_m$ and a graph $(T'_m,E)$ such that:
  \begin{itemize}
    \item $[z,z] \in T'_m$ for all $z$;
    \item any interval $[x,y] \in T_m$ is the union of no more than two intervals in $T'_m$; and
    \item for every $[x,y] \in T'_m$ and every $z \in [x,y]$, there exists a path in the graph $(T'_m,E)$ from $[x,y]$ to $[z,z]$.
  \end{itemize}
Then if a user is assigned to interval $[x,y]$, we know that $[x,y]$ is the union of no more than two intervals in $T'_m$ and for any $z \in [x,y]$ there is a path to $[z,z]$ in $(T'_m,E)$.
In other words, providing the user with the keys for the two appropriate intervals enables the user to derive all keys for which she is authorized.

We first observe that any interval $[x,y]$ such that $x \leqslant \ceil{m/2}$ and $y > \ceil{m/2}$ can be written as $[x,\ceil{m/2} \cup [\ceil{m/2}+1,y]$.
Recall that the binary decomposition construction splits $T_m$ in precisely this sort of way.
These observations suggest the following recursive construction.
For simplicity, we assume that $m$ is a power of two.

\begin{tissec}
\begin{Sch}[($2$-key binary decomposition)]\label{sch:2-key-construction}
We first apply Construction~\ref{sch:log-m-derivation} to $T_m$ to obtain a set of edges $E$.
We then identify the set of special nodes.
  \begin{enumerate}[$(1)$]
      \item If $m = 1$, then mark the node as a special node.
      \item If $m > 1$, then mark every node of the form
        \begin{enumerate}
          \item $[x,m/2]$, for $x < m/2$, as a special node;
          \item $[m/2+1,y]$, for $y > m/2 + 1$ as a special node.
        \end{enumerate}
      \item Split $T_m$ into $D_{m/2}$, $T^{\rm left}_{m/2}$ and $T^{\rm right}_{m/2}$ and recursively apply the node marking to $T^{\rm left}_{m/2}$ and $T^{\rm right}_{m/2}$.
  \end{enumerate}
Then $T'_m$ is defined to be the set of special nodes and we define the edge set to be $E' = E \cap \set{(x,y) : x,y \in T'_m}$.
\end{Sch}
\end{tissec}

\begin{arxiv}
\begin{Sch}[$2$-key binary decomposition]\label{sch:2-key-construction}
We first apply Construction~\ref{sch:log-m-derivation} to $T_m$ to obtain a set of edges $E$.
We then identify the set of special nodes.
  \begin{enumerate}[$(1)$]
      \item If $m = 1$, then mark the node as a special node.
      \item If $m > 1$, then mark every node of the form
        \begin{enumerate}
          \item $[x,m/2]$, for $x < m/2$, as a special node;
          \item $[m/2+1,y]$, for $y > m/2 + 1$ as a special node.
        \end{enumerate}
      \item Split $T_m$ into $D_{m/2}$, $T^{\rm left}_{m/2}$ and $T^{\rm right}_{m/2}$ and recursively apply the node marking to $T^{\rm left}_{m/2}$ and $T^{\rm right}_{m/2}$.
  \end{enumerate}
Then $T'_m$ is defined to be the set of special nodes and we define the edge set to be $E' = E \cap \set{(x,y) : x,y \in T'_m}$.
\end{Sch}
\end{arxiv}

\begin{figure}[h] \centering
  \includegraphics[width=\textwidth]{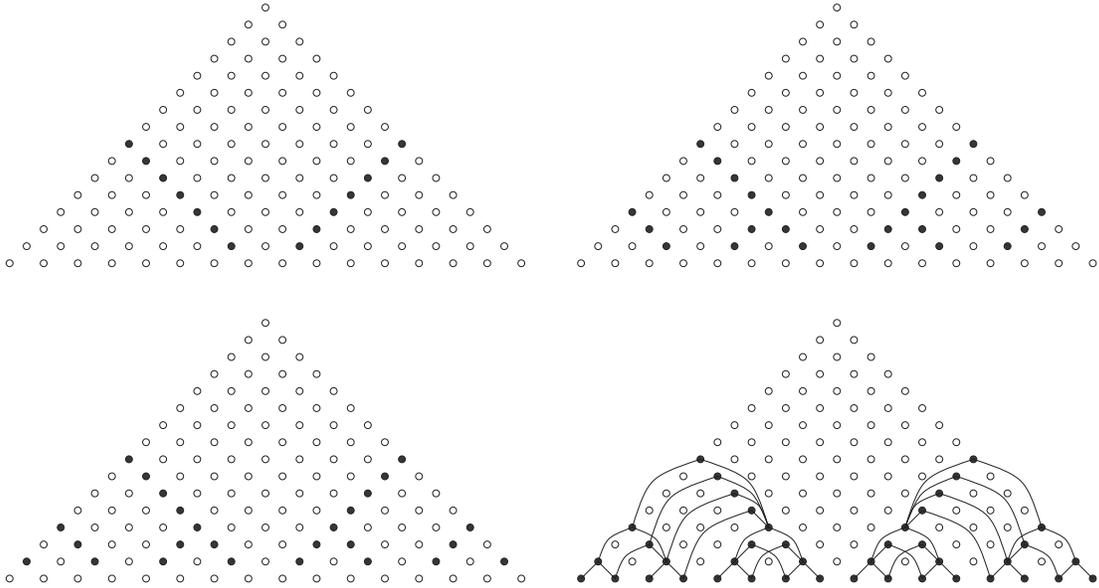}
\caption{Applying Construction~\ref{sch:2-key-construction} to $T_{16}$} \label{fig:2key-scheme-t16}
\end{figure}

In other words, $(T'_m,E')$ is the sub-graph of the binary decomposition of $T_m$ induced by the set of special nodes $T'_m$.
The result of applying $2$-key binary decomposition to $T_{16}$ is shown in Figure~\ref{fig:2key-scheme-t16}, in which the special nodes are represented by filled circles.
(The marking of leaf nodes and identification of edges are conflated in the final part of the figure.)

Then a key assignment scheme in which users may have two keys is implemented by constructing a key assignment scheme for the key derivation graph produced by Construction~\ref{sch:2-key-construction}.
A user associated with interval $[3,14]$, for example, would then be given the keys for intervals $[3,8]$ and $[9,14]$.

\begin{Pro}
Let $m = 2^d$ for some integer $d$.
Then there exists a set of enforcing edges $E$ such that $\card{E} < 2m\log m$, $(T_m,E)$ comprises two disconnected components, and the diameter of each component is $\log m/2$.
\end{Pro}

\begin{proof}
Clearly Construction~\ref{sch:2-key-construction} terminates after $\log m$ rounds.
Moreover, $s(m)$, the number of special nodes in $T_m$, satisfies the recurrence $s(m) < m + 2s(m/2)$, from which we deduce that $s(m) < m\log m$ and, since the out-degree of each special node equals $2$, we deduce that $\card{E} < 2m\log m$.
Clearly, the diameter of $(T_m,E)$ equals $\log m - 1  = \log m/2$.
\end{proof}

We also note that we can use a similar method to create a one-hop two-key scheme with $\bigO{n^2}$ edges.
In particular, we have the following result.

\begin{Pro}
There exists a set of enforcing edges $E$ such that $\card{E} < \frac{1}{2}m(m-1+\log m)$, $(T_m,E)$ comprises two disconnected components, and the diameter of each component is $1$.
\end{Pro}

\begin{proof}
Consider the set of edges $E$ of cardinality $\frac{1}{6}m(m-1)(m+4)$ that defines a $1$-hop graph for $T_m$ and then take the subgraph of $(T_m,E)$ induced by the set of special nodes.
From the top-left corner of Figure 5, it is evident that the number of edges in this subgraph $e(m)$ satisfies the following inequality:
  \begin{align*}
    e(2m) &< 2t_{m} + 2e(m), \\
  \intertext{where $t_{m}$ denotes the $m$th triangle number.  Hence, we have}
    e(2m) &< \frac{1}{2}m(m+1) + 2e(m) ,
  \end{align*}
from which we may prove the required result using a straightforward induction.
\end{proof}

\subsection{Related work}

A number of authors have used `binary tree encryption'', which can be used to enforce temporal access control policies cryptographically (see~\cite{back:secu06,cane:forw07}, for example).
In such schemes, assuming $n = 2^m$ for some integer $m$, the parent node is $[1,n]$ and the two child nodes of node $[x,y]$ are $[x,\frac{x+y}{2}]$ and $[\frac{x+y}{2}+1,y]$.
Then any interval $[x,y] \in T_n$ is the disjoint union of no more than $2(\log n-1)$ intervals: $[2,15] \in T_{16}$, for example, is the union of the intervals $[2,2]$, $[3,4]$, $[5,8]$, $[9,12]$, $[13,14]$ and $[15,15]$.
In other words, we may enforce a temporal access control policy using binary tree encryption, by supplying each user with at most $2(\log n - 1)$ keys and key derivation time requires no more than $\log n$ steps.
The great advantage of such schemes is that the key of a child node can be derived directly from the key of the parent, since the key derivation graph is a tree.
Hence, binary tree encryption schemes require no public information.

The focus of our work, however, is on schemes~--~arguably, more practical schemes~--~in which users have a single key.
Two groups of researchers have studied this form of cryptographic enforcement of temporal access control in some detail.
In this section, we discuss these two strands of research and our own work in this area.
We then summarize and compare the respective results to our work.

\begin{itemize}
  \item \citeN{atal:inco07} propose a number of schemes that reduce the number of edges and the maximum number of hops for $T_m$, using techniques previously developed for total orders~\cite{atal:key06}.
        Henceforth, we will use the term \emph{chain}, rather than total order, and we will write $C_n$ to denote the chain containing $n$ elements.

        \citeN{atal:inco07} treat $T_m$ as the direct product of two orthogonal sets of chains and apply short-cutting techniques for chains.
        To derive the key for interval $[z,z] \subseteq [x,y]$ given the key for $[x,y]$, for example, their method requires us to first derive the key for $[z,y]$ using one chain (comprising intervals of the form $[1,y],\dots,[x,y],\dots,[y,y]$), and then to derive the key for $[z,z]$ using an orthogonal chain (comprising intervals of the form $[z,m],\dots,[z,y],\dots,[z,z]$).

        Any set of enforcing edges $E$ for chain $C_m$, such that $\card{E} = \bigO{f(m)}$ and the diameter of the graph $(C_m,E)$ is $\bigO{d(m)}$, can be used to construct a set of enforcing edges $E'$ for $T_m$ such that $\card{E'}$ is $\bigO{m f(m)}$ and the diameter of $(T_m,E')$ is $\bigO{d(m)}$.
        More specifically, if the diameter of $(C_m,E)$ is $d$, then the diameter of $(T_m,E')$ is $2d$, and if $\card{E} = f(m)$, then $\card{E'} = 2\sum_{i=1}^m f(i)$.
  \item \citeN{desa:new08} propose a number of schemes that take a quite different approach, using earlier work due to \citeN{thor:shor95} and \citeN{dush:part41} to reduce the diameter of $T_m$, and to \citeN{alon:opti87} to reduce the number of edges (by increasing the number of keys).

        \citeN{thor:shor95} showed that given a rooted, acyclic, planar, directed graph \mbox{$G = (V,E)$}, there exists a set $E_s$ of ``shortcut edges'' such that $\card{E_s} \leqslant \card{E}$ and the diameter of $(V,E \cup E_s)$ is $\bigO{\log \card{V} \log^*\card{V}}$.

        The second result they use is that given a poset of dimension $2$ with Hasse diagram $G = (V,E)$, there exists a set of shortcut edges $E_s$ such that the graph $(V,E \cup E_s)$ has diameter $\bigO{\log^* \card{V}}$ and $\card{E_s}$ is $\bigO{d\card{V}(3\log \card{V})^{d-1}}$~\cite{desa:effi07,dush:part41}.

        \citeN{alon:opti87}, \citeN{yao:spac82}, and \citeN{bodl:trad94} have all studied \emph{$c$-coverings}, which are used to represent an interval as the union of no more than $c$ smaller intervals.
        \citeN{desa:new08} have considered the use of $c$-coverings in their work to create multi-key schemes in which each user requires no more than $c$ keys.
\end{itemize}
Both strands of research include enforcement schemes in which users may have multiple keys, as well as schemes in which users have precisely one key.
De Santis \etal\ provide several different ways of constructing key assignment schemes for $T_m$, whereas Atallah \etal\ rely on schemes for chains to construct their key assignment schemes.
The schemes of De Santis \etal\ provide greater flexibility in the choice of parameters, allowing, for example, a choice in the number of keys for a particular scheme.
In contrast, the schemes of Atallah \etal\ use three keys, because of the data structures that are used to build their schemes.
Moreover, the $3$-key schemes of Atallah \etal\ are rather artificial, in the sense that \emph{enabling keys} have to be introduced and more information needs to be stored both at the server side and the client side to enable key derivation to take place~\cite[\S 5.3 and 5.4]{atal:effi07}.

We previously introduced a number of schemes~\cite{cram:nordsec09}, including binary decomposition, which we generalize in this paper.
The main distinguishing feature of our work is the focus on improved schemes that are directly relevant to the problem at hand, whereas prior work has simply applied existing short-cutting techniques, without considering the particular characteristics of the graph $T_m$ and its application to temporal access control.
Specifically, in our previous work and in this paper we exploit the fact that it is not necessary to be able to derive keys for non-leaf intervals, in contrast to the work of other researchers.

As a consequence of our more direct approach, we are able to define schemes for which it possible to compute either exact values or tight upper bounds on storage and derivation costs, whereas related work only describes asymptotic behavior.
For large values of $m$, such a description may be useful, but, for smaller (and arguably more relevant) values of $m$, our approach is more informative.
Moreover, without knowing the multiplicative constants and lower-order terms hidden by the $\mathcal{O}$ notation, it is difficult to ascertain which scheme in the literature is the best to use for a particular value of $m$.
The relevant characteristics of comparable schemes in the literature and those introduced in this paper are summarized in Figure~\ref{tbl:comparison}.

\begin{figure}[!th]
  \subfigure[Single-key constructions]{
    \begin{minipage}{\textwidth}
      \[
        \begin{array}{|r|r|r|}
        \hline
          \textbf{Scheme} &\textbf{Public Storage} & \textbf{Derivation} \\
        \hline
          \text{\citeN[\S 4]{atal:inco07}} & \bigO{m^2 \log m} & 4 \\
                          & \bigO{m^2} &  \bigO{\log^* m} \\
        \hline
          \text{\citeN[\S 3.1]{desa:new08}} & \bigO{m^2} & \bigO{\log m \log^* m} \\ 
                           & \bigO{m^2 \log m} & \bigO{\log^* m} \\ 
                           & \bigO{m^2 \log m \log \log m} & 3 \\ 
        \hline
          \text{\citeN[\S 3.1]{cram:nordsec09}}  & m(m-1) & \ceil{\log m} \\
        \hline
          \text{\citeN[\S 5]{cram:nordsec09}}  & \frac{1}{6}m(m-1)(\sqrt{m} + 4) & 2 \\
        \hline
          \text{\bf This paper}  & m^2\left(1+ \frac{1}{6}\ceil{\log\log m}\right) & \ceil{\log\log m} \\
        \hline
        \end{array}
      \]
    \end{minipage}}
  \subfigure[Multi-key constructions]{
    \begin{minipage}{\textwidth}
      \[
        \begin{array}{|r|r|r|r|}
        \hline
          \textbf{Scheme} & \textbf{Keys} & \textbf{Public Storage} & \textbf{Derivation} \\
        \hline
          \text{Binary encryption trees} & 2(\log n - 1) & 0 & \log n \\
        \hline
          \text{\citeN[\S 5]{atal:inco07}}  & 3 & \bigO{m \log \log m} & 5 \\
                          & 3 & \bigO{m \log \log m} &  \bigO{\log^* m} \\
        \hline
          \text{\citeN[\S 4]{desa:new08}} & 2 & \bigO{m \log m} & \bigO{m} \\ 
                         & 3 & \bigO{m \log\log m} & \bigO{\sqrt{m}} \\ 
                         & 4 & \bigO{m \log^* m} & \bigO{\displaystyle\frac{m}{\log m}} \\ 
        \hline
          \text{\citeN[\S 4]{desa:new08}} & 2 & \bigO{m^2 \log m} & 1 \\ %
                         & 3 & \bigO{m \sqrt{m} \log\log m} & 1 \\ %
                         & 4 & \bigO{\displaystyle\frac{m^2}{\log m}} & 1 \\ 
        \hline
          \text{\citeN[\S 4]{desa:new08}} & \bigO{\log^*m} & \bigO{m} & \bigO{\displaystyle\frac{m}{(\log^* m)^3 \log m}} \\ %
                         & \bigO{\log^*m} & \bigO{m (\log^*m)^3} & 1 \\ 
        \hline
          \text{\bf This paper} & 2 & 2m\ceil{\log m} & \floor{\log m} \\
                          & 3 & 5m\ceil{\log \log m} & \log m + \ceil{\log\log m} \\
                          & 4 & 6m\ceil{\log^* m} & \log m + \ceil{\log^* m} \\
        \hline
        \end{array}
      \]
    \end{minipage}}
\caption{A comparison of existing work and our contributions} \label{tbl:comparison}
\end{figure}
%

%% file: geo_spatial3.tex
\section{Geo-Spatial Access Control}\label{sec:geo-spatial}

Atallah \etal\ have also applied their techniques to a graph in which the nodes correspond to rectangles of the form $I_1 \times I_2$, where $I_1 \in T_m$ and $I_2 \in T_n$~\cite{atal:effi07}.
Each object is associated with a point $(x,y)$ (equivalently, a ``unit'' rectangle  $[x,x]\times [y,y]$), where $x \in [1,m]$ and $y \in [1,n]$.
If a user is associated with node $I_1 \times I_2$ then the user should be able to derive the key for each point $[x,x] \times [y,y] \in I_1 \times I_2$ (where $x \in I_1$ and $y \in I_2$).
The set of points enclosed by a rectangle is defined by the endpoints of the two intervals.

There are $\frac{1}{2}m(m+1)$ and $\frac{1}{2}n(n+1)$ such intervals in $T_m$ and $T_n$, respectively, so there are a total of $\frac{1}{4}mn(m+1)(n+1)$ possible rectangles contained in $T_m \times T_n$ (and hence nodes in the graph).
Again, it is important to note that the user is not required to be able to derive keys for all subsets of $I_1 \times I_2$, only those subsets of the form $[x,x] \times [y,y]$.

Recall the binary decomposition technique for $T_m$ (Construction~\ref{sch:log-m-derivation}): for a given interval $[x,y]$ such that $x \leqslant \floor{m/2}$ and $y \geqslant \floor{m/2} + 1$, we connected $[x,y]$ to two intervals $[x,\floor{m/2}]$ and $[\floor{m/2}+1,y]$.
We now demonstrate how this technique can be extended to $T_{m,n}$.
Suppose, for illustrative purposes, that $m = n = 16$ and consider the rectangle $[3,11] \times [2,14]$ (as illustrated schematically in Figure~\ref{fig:rectangle-types}(a)).
Then we can decompose this rectangle into four smaller rectangles in which each interval contains no more than $8$ points: namely
  \[
    [3,11] \times [2,14] = ([3,8] \times [2,8]) \cup ([3,8] \times [9,14]) \cup ([9,11] \times [2,8]) \cup ([9,11] \times [9,14])
  \]
We can repeat this decomposition for each of these four rectangles, so that each interval contains no more than $4$ points.
It is easy to see that the out-degree of each node in the resulting graph can be no greater than $4$ and that the number of decompositions required (and hence the diameter of the resulting graph) is \mbox{$4 = \log 16$}.
Hence, we can construct an enforcing set of edges $E$ whose cardinality is bounded by $4\card{T_{n,n}} = n^2(n+1)^2$ and for which the diameter of the graph $(T_{n,n},E)$ is $\ceil{\log n}$.
However, we can reduce the number of edges by conducting a more detailed analysis.

In the next section, we provide a tighter bound on the number of edges required to construct a graph of diameter $\log n$ for $T_{n,n}$.
We then consider constructions for $T_{m,n}$, where $m \ne n$, and briefly look at multi-key constructions, before comparing our contributions with existing work.

\subsection{Constructions for $T_{n,n}$}

We first determine the number of edges required by a $1$-hop scheme.

\begin{Pro}
There exists an enforcing set of edges $E$ such that
  \[
    \card{E} < \frac{1}{36}n^2(n+1)^2(n+2)^2 = \frac{1}{9}(n+2)^2\card{T_{n,n}}
  \]
and the diameter of $(T_{n,n},E)$ is $1$.
\end{Pro}

\begin{proof}
Every node in $T_{n,n}$ is a rectangle and every point contained in each rectangle should be reachable in a single hop.
The area of each node corresponds to the number of leaf nodes that are reachable from this node.
Hence, the number of edges $e(n)$ required to construct a graph of diameter $1$ for $T_{n,n}$ is $\sum_{i=1}^n \sum_{j=1}^n a_{ij} ij$, where $a_{ij}$ is the number of rectangles of area $ij$.
Now the number of rectangles of area $ij$ is the number of intervals of length $i$ multiplied by the number of intervals of length $j$.
Hence, we have
  \[
    e(n) = \sum_{i=1}^n\sum_{j=1}^n (n-i+1)(n-j+1)ij = \left(\sum_{i=1}^n (n-i+1)i\right)^2 = \frac{1}{36}n^2(n+1)^2(n+2)^2.
  \]
In fact, $e(n)$ is an overestimate for the number of edges required because we have included $n^2$ rectangles of area $1$, which are leaf nodes.
The result follows.
\end{proof}



\begin{Thm}\label{thm:square-grid}
There exists an enforcing set of edges $E$ such that
  \[
    \card{E} = \frac{1}{3}n^2(n-1)(2n+5) < \frac{8}{3}\card{T_{n,n}} 
  \]
and the diameter of the graph $(T_{n,n},E)$ is bounded by $\ceil{\log n}$.
\end{Thm}

\begin{proof}
Given $n = 2m$, for some integer $m$, we can divide the $n \times n$ grid into four $m \times m$ grids, which we label $\reclabel{m}{0}{0}$, $\reclabel{m}{0}{1}$, $\reclabel{m}{1}{0}$, and $\reclabel{m}{1}{1}$.
Then there are three types of rectangles in $T_{n,n}$ (illustrated in Figure~\ref{fig:rectangle-types}):
  \begin{enumerate}
    \item those in which each vertex is in a different $m$-grid;
    \item those in which one pair of vertices is contained in one $m$-grid and the other pair of vertices are in an adjacent $m$-grid;
    \item those in which each vertex is in the same $m$-grid.
  \end{enumerate}

\begin{figure}[h]
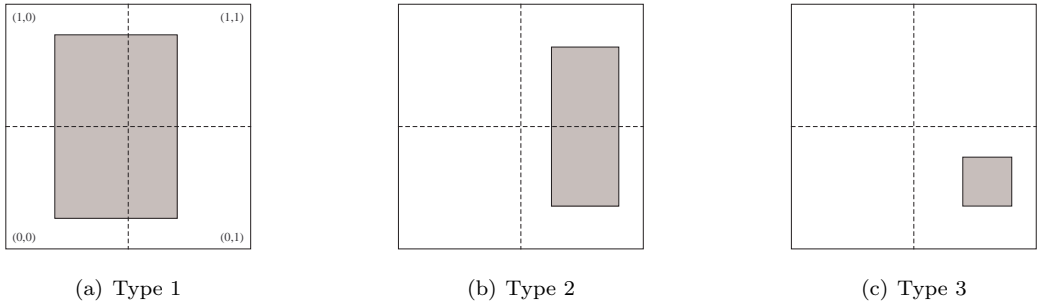
 \centering
  \subfigure[Type 1]{\begin{minipage}{.3\textwidth}\centering\includegraphics[width=.8\textwidth]{rectangles_type1.md}\end{minipage}}
  \hfill
  \subfigure[Type 2]{\begin{minipage}{.3\textwidth}\centering\includegraphics[width=.8\textwidth]{rectangles_type2.md}\end{minipage}}
  \hfill
  \subfigure[Type 3]{\begin{minipage}{.3\textwidth}\centering\includegraphics[width=.8\textwidth]{rectangles_type3.md}\end{minipage}}
\caption{Examples of rectangle types in a $2m \times 2m$ grid} \label{fig:rectangle-types}
\end{figure}

Then we construct an edge set in which:
  \begin{itemize}
    \item each Type 1 rectangle is connected to four child rectangles, one in each $m$-grid; and
    \item each Type 2 rectangle is connected to two child rectangles, one in each $m$-grid that contains a pair of the rectangle's vertices.
  \end{itemize}
We then recursively construct an edge set for each copy of $T_{m,m}$.
Hence, the number of edges $e(n)$ required by this construction is $4a + 2b + 4e(m)$, where $a$ represents the number of Type 1 rectangles and $b$ represents the number of Type 2 rectangles.

We now compute $a$ and $b$.
Note that we have $m$ choices for each of the four endpoints of the intervals that define a Type 1 rectangle (since each vertex lies in a different $m$-grid).
Hence, $a = m^4$.
To compute $b$, we first consider the number of rectangles in a pair of adjacent $m$-grids: we have $m^2$ choices for the endpoints of the ``long'' side of a Type 2 rectangle (that spans two $m$-grids), and $\frac{1}{2}m(m+1)$ choices for the endpoints of the ``short'' side (where the endpoints belong to the same $m$-grid).
Clearly, there are four different choices of adjacent $m$-grids, each containing the same number of Type 2 rectangles.
Hence, $b = 2m^3(m+1)$.
Therefore, we deduce the following recurrence relation for $e(n)$.
  \begin{align*}
    e(n) &= 4m^4 + 4m^3(m+1) + 4e(m) \\
         &= 4m^3(2m+1) + 4e(m) \\
         &= \frac{1}{2}n^3(n+1) + 4e\left(\frac{n}{2}\right)
  \end{align*}
Using this recurrence relation, we prove by induction that $e(n) = \frac{1}{3}n^2(n-1)(2n+5)$.

Consider $n = 2$: it is easy to see by inspection that we require $4 + 8 = 12$ edges, so the result holds for $n=2$ (since, using the formula, $e(2) = \frac{1}{3} \cdot 4 \cdot 1 \cdot 9 = 12$).
Now assume that the result holds for all $n < N$.
Then
  \begin{align*}
    e(N) &= \frac{1}{2}N^3(N+1) + 4e\left(\frac{N}{2}\right) \qquad\text{(by the recurrence relation)} \\
          &= \frac{1}{2}N^3(N+1) + \frac{4}{3}\frac{N^2}{4}\left(\frac{N}{2}-1\right)(N+5) \qquad\text{(by the inductive hypothesis)} \\
          &= \frac{1}{6}N^2\left(3N^2 + 3N + N^2 + 3N - 10\right) \\
          &= \frac{1}{3}N^2(N-1)(2N+5)
  \end{align*}
as required.
Moreover,
  \[
    e(n) = \frac{1}{3}n^2(n-1)(2n+5) = \frac{1}{3}n^2(2n^2+3n-5) < \frac{2}{3}n^2(n+1)^2 = \frac{8}{3}\card{T_{n,n}}.
  \]
It is evident that the construction terminates after no more than $\ceil{\log n}$ iterations and that the diameter of the resulting graph will be $\ceil{\log n}$.
\end{proof}

\subsection{Constructions for $T_{m,km}$}

We now consider constructions for $T_{m,km}$, where $m$ and $k$ are integers.
(These constructions can be extended to $T_{m,n}$, for any integers $m \leqslant n$, by writing $n = km + r$, where $0 \leqslant r < m$.)

\begin{figure}[h] \centering
  \includegraphics[scale=1.2]{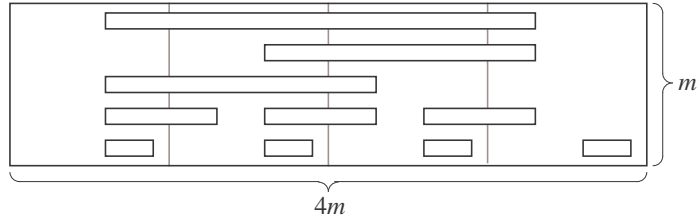}
\caption{A $m \times 4m$ grid and the various types of rectangles that may arise in it} \label{fig:m-km-grid}
\end{figure}

We consider an $m \times km$ grid to be $k$ copies of an $m \times m$ grid, which we may label $\reclabel{m}{1}{1},\dots,\reclabel{m}{1}{k}$.
Then all four vertices of a rectangle in $T_{m,km}$ may belong to the same $m$-grid, or one pair of vertices belongs to one $m$-grid and the other pair to another grid.
In other words, the number of possible choices of $m$-grids for the vertices of a rectangle corresponds to the number of intervals in $[1,k]$.
This is illustrated schematically for a $m \times 4m$ grid in Figure~\ref{fig:m-km-grid}.
These observations suggest the following approach:
  \begin{itemize}
    \item For each rectangle in $T_{m,km}$ that is not included in one copy of $T_{m,m}$, add edges to the appropriate rectangles in two or more copies of $T_{m,m}$;
    \item Construct an enforcing set of edges for each copy of $T_{m,m}$.
  \end{itemize}
In Section~\ref{sec:temporal} we identified a number of schemes for $T_k$ and we have seen (in Theorem~\ref{thm:square-grid}) how to construct an enforcing set of edges $E$ such that the diameter of $T_{m,m}$ is $\log m$.
Putting this together, we obtain the following result.

\begin{Thm}\label{thm:rectangular-grid}
There exist enforcing sets of edges $E_1$ and $E_2$ such that
  \begin{itemize}
    \item $\card{E_1} = \frac{1}{12}km^2\big((k-1)(k+4)m(m+1) + 4(m-1)(2m+5)\big)$ and the diameter of $(T_{m,km},E_1)$ is $\log m + 1 = \log 2m$;
    \item $\card{E_2} = \frac{1}{6}km^2\big(3(k-1)m(m+1) + 2(m-1)(2m+5)\big)$ and the diameter of $(T_{m,km},E_2)$ is $\log m + \log k = \log km$.
  \end{itemize}
\end{Thm}

\begin{proof}
For each rectangle that belongs to two or more $m$-grids we have $m^2$ choices for the vertices that belong to different $m$-grids  and $\frac{1}{2}m(m+1)$ choices for those that belong to the same $m$-grid.
Hence, in total we have $\frac{1}{2}m^3(m+1)$ possible choices for such rectangles.
Hence, we can build a set of enforcing edges for the $m \times km$ grid, whose cardinality is given by
  \begin{equation}\label{eq:edges-t-m-km}
    \frac{1}{2}m^3(m+1)e(T_k) + ke(T_{m,m}),
  \end{equation}
where $e(T_k)$ is the cardinality of some enforcing set of edges for $T_k$ and $e(T_{m,m})$ is the cardinality of some enforcing set of edges for $T_{m,m}$.
The number of hops required will be the number of hops for $T_k$ plus the number of hops required for $T_{m,m}$.

Each rectangle in $T_{m,km}$ has non-empty intersection with $\reclabel{m}{1}{x},\dots,\reclabel{m}{1}{y}$ for some $x$ and $y$.
To obtain $E_1$, we add an edge from each rectangle in $T_{m,km}$ to a single rectangle in each of $\reclabel{m}{1}{z}$, $z \in [x,y]$.
In other words, the number of edges required will be the number of edges required for a $1$-hop scheme for $T_k$ multiplied by the number of rectangles in each interval.
Hence, using the $1$-hop construction in the proof of Theorem~\ref{thm:multiplicative-decomposition-repeated-factors} and applying~\eqref{eq:edges-t-m-km}, we have
  \begin{align*}
    \card{E_1} &= \frac{1}{12}m^3(m+1)k(k-1)(k+4) + \frac{1}{3}km^2(m-1)(2m+5) \\
               &= \frac{1}{12}km^2\big((k-1)(k+4)m(m+1) + 4(m-1)(2m+5)\big)
  \end{align*}
We require a single hop to get from any rectangle to a rectangle in a copy of $T_{m,m}$ and we require $\log m$ hops to get from any rectangle in $T_{m,m}$  to a leaf node.
Hence, the diameter of $(T_{m,km},E_1)$ is $1 + \log m$.

To obtain $E_2$, we construct an enforcing set of edges that enables us to get from each rectangle in $T_{m,km}$ to a rectangle in a copy of $T_{m,m}$ in $\log k$ hops.
Using Construction~\ref{sch:log-m-derivation}, we require $k(k-1)$ edges for each rectangle.
Hence,
  \begin{align*}
    \card{E_2} &= \frac{1}{2}m^3(m+1)k(k-1) + \frac{1}{3}km^2(m-1)(2m+5) \\
               &= \frac{1}{6}km^2\big(3(k-1)m(m+1) + 2(m-1)(2m+5)\big)
  \end{align*}
In this case, the number of hops required in total is $\log k + \log m$.
\end{proof}

\begin{Cor} \label{cor:geo-spatial-bound}
For $k \geqslant 1$, there exists an enforcing set of edges $E$ such that
  \[
    \card{E} < 2\card{T_{m,km}}\left(1+\frac{1}{3k}\right) \leqslant \frac{8}{3}\card{T_{m,km}}
  \]
and the diameter of $(T_{m,km},E)$ is $\log km$.
\end{Cor}

\begin{proof}
The result for $k=1$ follows from Theorem~\ref{thm:square-grid}.
For $k > 1$, we have (from the proof of Theorem~\ref{thm:rectangular-grid})
  \begin{align*}
    E_2 &= \frac{1}{6}km^2\big(3(k-1)m(m+1) + 2(m-1)(2m+5)\big) \\
        &= \frac{1}{6}km^2(3km^2+3km-3m^2-3m+4m^2+6m-10) \\
        &= \frac{1}{6}km^2(3km(m+1)+m^2+3m-10) \\
        &< \frac{1}{6}km^2(3km(m+1)+(m+1)(m+2)) \\
        &= \frac{1}{6}km^2(3km+m+2)(m+1) \\
        &= \frac{1}{6}m(m+1)km(3km+3+m-1) \\
        &= \frac{1}{2}m(m+1)km(km+1)+\frac{1}{6}m(m+1)km(m-1)
    \intertext{and, since $\card{T_{m,km}} = \frac{1}{4}m(m+1)km(km+1)$, we have}
    E_2 &= 2\card{T_{m,km}} + \frac{2}{3}\left(\frac{m-1}{km+1}\right)\card{T_{m,km}} < 2\card{T_{m,km}}\left(1+\frac{1}{3k}\right)
  \end{align*}
as required.
Clearly $1 + \frac{1}{3k}$ monotonically decreases as $k$ increases, so $\card{E_2}$ is a maximum when $k = 1$.
Hence $\card{E_2} < \frac{8}{3}\card{T_{m,km}}$.
\end{proof}

\subsection{Multi-key constructions}\label{sec:geo-spatial-multi}

In Section~\ref{sec:temporal-multi-key}, we showed how we could reduce the number of edges in an enforcing set if we assumed that a user may be given two keys.
Essentially, this assumption allows us to reduce the number of nodes in the key derivation graph for $T_n$ from approximately $\frac{1}{2}n^2$ to $n\log n$.
We now develop an analogous approach for $T_{n,n}$.

\begin{figure}[h]
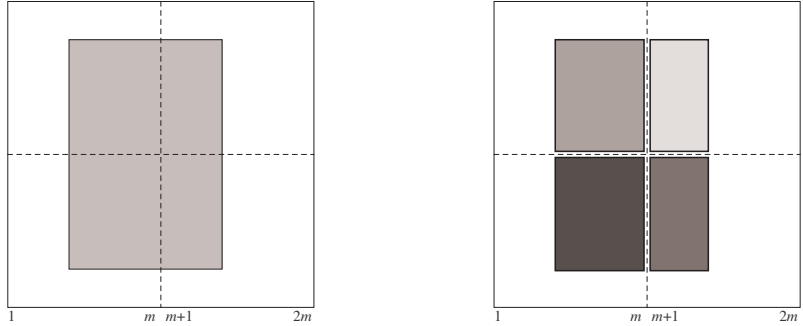

    \hspace*{\fill}
  \subfigure{\begin{minipage}{.3\textwidth}\centering\includegraphics[width=\textwidth]{special_rectangles0.md}\end{minipage}}
    \hfill
  \subfigure{\begin{minipage}{.3\textwidth}\centering\includegraphics[width=\textwidth]{special_rectangles.md}\end{minipage}}
    \hspace*{\fill}
\caption{Representing a Type 1 rectangle in $T_{2m,2m}$ as the union of four special rectangles}\label{fig:type1-rectangle-as-4-special-rectangles}
\end{figure}

First, we explain how the set of special rectangles is defined.
We divide $T_{n,n}$ into four copies of $T_{m,m}$, where $m = n/2$.
Then the following rectangles are defined to be special nodes:
  \begin{itemize}
    \item $[x,m] \times [y,z]$, where $x,y,z \in [1,m-1]$;
    \item $[x,y] \times [z,m]$, where $x,y,z \in [1,m-1]$;
    \item $[m+1,x] \times [y,z]$, where $x \in [m+2,2m]$, $y,z \in [1,m-1]$;
    \item $[x,y] \times [z,m]$, where $x,y \in [m+2,2m]$, $z \in [1,m-1]$;
    \item $[x,m] \times [y,z]$, where $x \in [1,m-1]$, $y,z \in [m+2,2m]$;
    \item $[x,y] \times [m+1,z]$, where $x,y \in [1,m-1]$, $z \in [m+2,2m]$;
    \item $[m+1,x] \times [y,z]$, where $x \in [m+2,2m]$, $y,z \in [m+2,2m]$;
    \item $[x,y] \times [m+1,z]$, where $x,y \in [m+2,2m]$, $z \in [m+2,2m]$.
  \end{itemize}
In other words, special nodes in $T_n$, where $n = 2m > 2$, are non-leaf rectangles in which at least one endpoint is $m$ or $m+1$.
Having identified the special nodes in $T_{n,n}$, we then recursively identify the special nodes in each copy of $T_{m,m}$.
Figure~\ref{fig:type1-rectangle-as-4-special-rectangles} illustrates how a Type 1 rectangle can be split into four special rectangles.

\begin{Thm}
Let $n = 2m$ for some integer $m$.
Then there exists an enforcing set of edges $E$ for $T_{n,n}$ such that $\card{E} \leqslant 4n^2(n-1)$, the graph $(T_{n,n},E)$ comprises four disconnected components, and the diameter of $(T_{n,n},E)$ is $\floor{\log n}$.
\end{Thm}

\begin{proof}
We first count the number of special rectangles.
Without loss of generality, we consider the copy of $T_{m,m}$ in which all special rectangles contain an interval of the form $[x,m]$ for some $x \in [1,m-1]$ (corresponding to the bottom left quadrant in Figure~\ref{fig:type1-rectangle-as-4-special-rectangles}).%
  \footnote{By symmetry, each copy of $T_{m,m}$ contains the same number of special rectangles.}
Then the number of special nodes is the total number of rectangles in $T_{m,m}$ minus the number of rectangles that are not special nodes.
Since a non-special rectangle cannot contain an interval in which the upper endpoint is equal to $m$, the number of non-special rectangles is $\frac{1}{4}(m-1)^2 m^2$.
Hence, the number of special rectangles is given by
  \[
    \frac{1}{4}m^2(m+1)^2 - \frac{1}{4}m^2(m-1)^2 = m^3.
  \]
By symmetry, $T_{n,n}$ contains $4m^3 = \frac{1}{2}n^3$ special rectangles in which at least one endpoint is $m$ or $m+1$.

The recursive construction implies that $s(n)$, the total number of rectangles that will be marked as special, satisfies the recurrence $s(n) \leqslant n^3/2 + 4s(n/2)$, from which we deduce that $s(n) \leqslant n^2(n-1)$.
Since each special rectangle has out-degree no greater than $4$, we conclude that there exists an enforcing set of edges of cardinality no greater than $4n^2(n-1)$.
\end{proof}

\subsection{Related work}

\citeN[\S 4]{atal:effi07} propose a scheme for geo-spatial access control in which each user has a single key, the number of key derivation steps is $\bigO{1}$ and the number of edges is $\bigO{n^4 (\log n^2)^3 \log^* n}$.
They then propose more complex schemes in which the user has $\bigO{1}$ keys.
These schemes require complex, auxiliary data structures and key derivation algorithms.
The number of edges required by their best scheme for $T_{m,n}$, where $m \geqslant n$, is $\bigO{mn(\log\log m)^2 \log^* m}$.

It is difficult to compare the performance of our schemes with those of Atallah \etal\ because most of their schemes use $\bigO{1}$ keys.
The only scheme that uses a single key has constant time key derivation (requiring no more than $9$ hops) 
and requires $\bigO{m^2n^2(\log mn)^3 \log^* mn}$ edges.
The scheme is rather complicated and involves reducing the $4$-dimensional poset $(T_{m,n},\subseteq)$ to a set of $1$-dimensional posets (that is, chains) and then constructing edge sets for each of these chains and edge sets to connect the chains.

Recently, \citeN{yuan:effi09} used $\bigO{mn}$ copies of a scheme for a chain to create a scheme in which the user requires $\bigO{1}$ keys and $\bigO{mn\log^* m}$ edges, and the number of derivation steps is $\bigO{1}$.
However, the paper is rather vague on the details of the scheme and how it compares to the closely related work of \citeN{atal:effi07} described in the preceding paragraphs, so it is difficult to provide a direct comparison with our work.

By Corollary~\ref{cor:geo-spatial-bound}, the number of edges required by our scheme is less than $\frac{2}{3}mn(m+1)(n+1)$ and key derivation takes no more than $\log{mn}$ steps.
Moreover, our construction is analogous to our binary decomposition for temporal access control: we will see in the next section that our construction generalizes readily to higher dimensions; in contrast, the extension of existing schemes to higher dimensions is non-trivial.

%% file: interval1.tex
\section{Interval-Based Access Control} \label{sec:interval}

In this section, we generalize temporal and geo-spatial access control to \emph{interval-based access control}.
Consider
  \[
    T_n^k \stackrel{\rm def}{=} \underset{\text{$k$ times}}{\underbrace{T_n \times \dots \times T_n}}.
  \]
We call an element $[x_1,y_1] \times \dots \times [x_k,y_k] \in T_n^k$ a ($k$-dimensional) \emph{hyperrectangle} and write it $\prod_{i=1}^k [x_i,y_i]$.
A $1$-dimensional hyperrectangle is an interval (as in Section~\ref{sec:temporal}) and a $2$-dimensional hyperrectangle is simply a rectangle (as in Section~\ref{sec:geo-spatial}).
In interval-based access control, protected objects are associated with a ``trivial'' hyperrectangle $\prod_{i=1}^k [x_i,x_i]$ (which is simply a point in $k$-dimensional space) and users are associated with a hyperrectangle $\prod_{i=1}^k [x_i,y_i]$.
A user associated with hyperrectangle $\prod_{i=1}^k [x_i,y_i]$ is authorized for an object associated with $\prod_{i=1}^k [z_i,z_i]$ if and only if $z_i \in [x_i,y_i]$ for all $i$.

\begin{Thm}\label{thm:generalization}
There exists a set of enforcing edges $E$ for $T_n^k$ such that
  \[
    \card{E} = \frac{n^k}{2^k}\sum_{i=1}^k \binom{k}{i}\frac{(3^i-1)(n^i-1)}{2^i-1}
  \]
and the diameter of $(T_n^k,E)$ is $\log n$.
\end{Thm}

Note that substituting $k=1$ and $k=2$ into the above formula, we obtain $\card{E} = n(n-1)$ and $\card{E} = \frac{1}{3}n^2(n-1)(2n+5)$, confirming the results of Sections~\ref{sec:temporal} and~\ref{sec:geo-spatial}.
Before proving the above theorem, we state a useful result, which can be proved by induction.

\begin{Pro} \label{pro:recurrence}
Let $k \geqslant 1$, $i \geqslant 0$ and $a_0,\dots,a_t$ be integers and let $n = 2^m$ for some non-zero positive integer $m$.
If
  \begin{align*}
    &f(n) - 2^k f\left(\frac{n}{2}\right) = \left(\frac{n}{2}\right)^k \sum_{i=0}^t a_i \left(\frac{n}{2}\right)^{i},
   \intertext{for all $n$, and $f(1) = 0$, then}
    &f(n) = \left(\frac{n}{2}\right)^k \left(a_0  \log n + \sum_{i=1}^t a_i \frac{n^i-1}{2^i-1}\right)
  \end{align*}
\end{Pro}

%

\begin{proof}[of Theorem~\ref{thm:generalization}]
Let $n = 2m$.
Then we may split $T_n^k$ into $2^k$ copies of $T_m^k$.
A hyperrectangle has non-empty intersection with one or more of the $2^k$ copies of $T_m^k$.
Our proof proceeds by counting the number of copies of $T_m^k$ with which a hyperrectangle intersects and how that, in turn, determines the number of edges required for that hyperrectangle.

We first recall the methods of Sections~\ref{sec:temporal} and~\ref{sec:geo-spatial}.
For $k = 1$ (Section~\ref{sec:temporal}) we split $T_{n}$ into two copies of $T_m$ and every interval in $T_n$ has non-empty intersection with either one or two copies of $T_m$.
For $k = 1$, the two copies of $T_m$ may be identified with the $1$-bit string $0$ and $1$.
The endpoints of $[x,y] \in T_n$ are simply $x$ and $y$.
Then either both endpoints (that is, $x$ and $y$) belong to the same copy of $T_m$ or they are in different copies.
In the first case, there are $\frac{1}{2}m(m+1)$ choices for the endpoints, since we require that $x \leqslant y$; in the second case, there are $m^2$ choices, since we choose any value from the copy of $T_m$ corresponding to $0$ for the lower endpoint and any value from the copy of $T_m$ corresponding to $1$ for the upper endpoint.
And when $k = 2$ (Section~\ref{sec:geo-spatial}), $T_n^2$ may be split into four copies of $T_m$ labeled $(0,0)$, $(0,1)$, $(1,0)$ and $(1,1)$.
An element of $T_n^2$ is a rectangle, which ``straddles'' one, two or four copies of $T_m$ depending on which of the four copies of $T_m$ contain the lower left and upper right corners of the rectangle (as we saw in the proof of Theorem~\ref{thm:square-grid}).

More generally, each copy of $T_m$ in $T_n^k$ can be identified with a $k$-bit string.
Then each element of $T_n^k$ has two $k$-dimensional endpoints, which uniquely identify the two copies of $T_m$ to which those endpoints belong.
More generally, each hyperrectangle in $T_n^k$ is enclosed by some hyperrectangle comprising $2^d$ copies of $T_m$ for some integer $d \leqslant k$, where $d$ is determined by the endpoints.
We wish to determine $d$ for a given element of $T_n^k$.

We denote the ``left'' or ``lower'' endpoint of an element of $T_n^k$ by $l = (l_1,\dots,l_k) \in \set{0,1}^k$ and the ``right'' or ``upper'' endpoint by $r = (r_1,\dots,r_k) \in \set{0,1}^k$.
Then we have $l_i \leqslant r_i$, for all $i$.
Now the Hamming distance of $l$ and $r$ determines the ``volume'' of the enclosing hyperrectangle, measured as multiples of $T_m^k$.
Specifically, if the Hamming distance between the two endpoint strings is $d$ (that is, $l_i = 0$ and $r_i = 1$ for $d$ values of $i$) then the enclosing hyperrectangle contains $2^d$ copies of $T_m$.

The number of pairs of strings with Hamming distance $d$ is determined by the choice of co-ordinates at which $l_i \ne r_i$ and the values chosen for the remaining positions.
Clearly there are $\binom{k}{d}$ choices of $d$ co-ordinate positions and, having fixed those positions, there are $2^{k-d}$ choices for the values of the remaining $k-d$ positions.
Hence, the number of different endpoints that are enclosed in a hyperrectangle of $2^d$ copies of $2^m$ is $2^{k-d}\binom{k}{d}$.

Now an arbitrary element in $\prod_{i=1}^k [x_i,y_i] \in T_n^k$ for which the Hamming distance of the endpoints is $d$ ``straddles'' $2^d$ copies of $T_m$, and is the union of $2^d$ elements of $T_m^k$.
(When $k = 1$, for example, every interval in $T_n$ is the union of $1$ or $2$ intervals contained in $\trlabel{m}{0}$ and $\trlabel{m}{1}$.)
Therefore, $2^d$ edges will be required to connect $\prod_{i=1}^k [x_i,y_i]$ to the appropriate child hyperrectangles that are contained in copies of $T_m$.

For an arbitrary element $\prod_{i=1}^k [x_i,y_i] \in T_n^k$ with endpoints $l,r \in \set{0,1}^k$, the value of $l_i \oplus r_i$ determines how many choices there are for $x_i$ and $y_i$.
Specifically, if $l_i = r_i$, then $x_i$ and $y_i$ belong to the same $m$-cube and there are $\frac{1}{2}m(m+1)$ choices for the pair $(x_i,y_i)$ since we must ensure that $y_i \geqslant x_i$.
However, if $l_i < r_i$, then $x_i$ and $y_i$ belong to different $m$-cubes and there are $m^2$ choices for $(x_i,y_i)$, since $y_i$ is necessarily greater than $x_i$ and therefore we have a free choice of $x_i$ and $y_i$ from $m$ values.
Therefore, if the Hamming distance of $l$ and $r$ is $d$, then the total number of choices for $\prod_{i=1}^k [x_i,y_i]$ is
  \[
    (m^2)^d\left(\frac{1}{2}m(m+1)\right)^{k-d} = \frac{1}{2^{k-d}}m^{k+d}(m+1)^{k-d}
  \]
We conclude that the set of edges required to connect all hyperrectangles with Hamming distance $d$ to the appropriate child hyperrectangles has cardinality $a(d) b(d) c(d)$, where
  \begin{itemize}
    \item $a(d) = 2^d$ is the number of children of a hyperrectangle with Hamming distance $d$,
    \item $b(d) = 2^{k-d}\binom{k}{d}$ is the number of possible choices of enclosing hyperrectangles for a hyperrectangle with Hamming distance $d$, and
    \item $c(d) = \frac{1}{2^{k-d}}m^{k+d}(m+1)^{k-d}$ is the number of choices of endpoints for each of the $k$ intervals comprising a hyperrectangle with Hamming distance $d$.
  \end{itemize}
Clearly, the Hamming distance $d$ takes values between $0$ and $k$, so the total number of edges required, denoted $e(n,k)$, satisfies the recurrence relation:
  \begin{align*}
    e(n,k) &= 2^k e(m,k) + \sum_{d=0}^k 2^d 2^{k-d}\binom{k}{d} \frac{1}{2^{k-d}}m^{k+d}(m+1)^{k-d} \\
           &= 2^k e(m,k) + m^k \sum_{d=0}^k \binom{k}{d} (2m)^d (m+1)^{k-d} \\
    \intertext{and, applying the binomial theorem, we obtain}
    e(n,k) &= 2^k e(m,k) + m^k (2m+m+1)^k = 2^k e(m,k) + m^k(3m+1)^k.
  \end{align*}
Now, this total includes an edge for each hyperrectangle that is contained in a single $m$-cube (when the Hamming distance is $0$).
For a recursive construction, we can omit these edges (of which there are $2^k \frac{1}{2^k}m^k(m+1)^k = m^k(m+1)^k$).
Hence, subtracting the edges for hyperrectangles with Hamming distance $0$, we obtain
  \begin{align*}
    e(n,k) &= 2^k e(m,k) + m^k(3m+1)^k - m^k(m+1)^k \\
           &= 2^k e(m,k) + m^k \left(\sum_{i=1}^k \binom{k}{i}(3m)^i - \sum_{i=1}^k \binom{k}{i} m^i\right) \\
           &= 2^k e(m,k) + m^k \sum_{i=1}^k \binom{k}{i}(3^i-1)m^i.
    \intertext{Replacing $m$ by $n/2$, we obtain}
    e(n,k) &= 2^k e\left(\frac{n}{2},k\right) + \sum_{i=1}^k \binom{k}{i}(3^i-1)\left(\frac{n}{2}\right)^{i+k}.
  \end{align*}
Moreover, $e(1,k) = 0$ for all $k$.
Hence, we may apply Proposition~\ref{pro:recurrence}, thereby obtaining
  \begin{equation} \label{eq:edges-general-case}
    e(n,k) = \frac{n^k}{2^k} \sum_{i=1}^k \binom{k}{i}\frac{(3^{i}-1)(n^i-1)}{2^i - 1}
  \end{equation}
as required.

Clearly the number of derivation steps $d(n)$ obeys the recurrence relation \mbox{$d(n) = 1 + d(n/2)$}, from which it immediately follows that $d(n) = \log n$.
\end{proof}

\begin{Cor}
There exists a set of enforcing edges $E$ such that
  \[
    \frac{(3^k-1)n^k(n^k-1)}{2^k(2^k-1)} < \card{E} < \frac{(3^k-1)n^k(n^k+1)}{2^k(2^k-1)}
  \]
and the diameter of $(T_n^k,E)$ is $\log n$.
\end{Cor}

\begin{proof}
The term in the left-hand side of the inequality is simply the last term in the summation in~\eqref{eq:edges-general-case}.
Now note that
  \[
    \frac{3^i - 1}{2^i-1} < \frac{3^{i+1}-1}{2^{i+1}-1}
  \]
for all $i \geqslant 1$.
Hence, we have
  \begin{align*}
    e(n,k) &= \frac{n^k}{2^k} \sum_{i=1}^k \binom{k}{i}\frac{(3^{i}-1)(n^i-1)}{2^i - 1} < \frac{n^k(3^k-1)}{2^k(2^k-1)}\sum_{i=1}^k \binom{k}{i}(n^i-1) \\
           &< \frac{n^k(3^k-1)}{2^k(2^k-1)}\sum_{i=1}^k \binom{k}{i}n^i < \frac{n^k(3^k-1)}{2^k(2^k-1)}(n+1)^k,
  \end{align*}
as required.
\end{proof}
\begin{Rem}
Since $\card{T_n^k} = \frac{1}{2^k}n^k(n+1)^k$, the above result implies that $\card{E}$ is $\Theta\left(\left(\frac{3}{2}\right)^k \card{T_n^k}\right)$.
\end{Rem}

To conclude this section, we state and prove a result that provides an upper bound on the number of edges required in an enforcing set when we assume that a user may have up to $2^k$ keys.
We do not describe the recursive procedure that is used to label special nodes, as it is a straightforward generalization of the techniques described in Sections~\ref{sec:temporal-multi-key} and~\ref{sec:geo-spatial-multi}.

\begin{Thm}
There exists a set of enforcing edges $E$ such that \[ \card{E} \leqslant 2kn^k(n^{k-1} + \log n - 1), \] the graph of $(T_n^k,E)$ comprises $2^k$ disconnected components and the diameter of the graph is $\log n$.
\end{Thm}

\begin{proof}
Let $n = 2m$ for some integer $m$.
Then we can divide the hypercube $T_n^k$ into $2^k$ copies of the hypercube $T_m^k$.
As before, without loss of generality, we consider the hypercube $T_m^k$ in which all intervals are of the form $[x_i,y_i]$ with $[x_i,y_i] \in [1,m]$.
The special hyperrectangles in this copy of $T_m^k$ have the form $\prod_{i=1}^k [x_i,y_i]$, where $x_i = m$ or $y_i = m$ for some $i$.
Since we can choose any one of $k$ intervals in which to fix an endpoint, the total number of special rectangles in a particular copy of $T_m^k$ is no greater than
  \[
    km\left(\frac{1}{2}m(m+1)\right)^{k-1} = \frac{1}{2^{k-1}}km^k(m+1)^{k-1}.
  \]
There are $2^k$ copies of $T_m^k$, hence the total number of special rectangles of the form $\prod_{i=1}^k [x_i,y_i]$, where $x_i = m+1$ or $y_i = m$ for some $i$, is $2km^k(m+1)^{k-1}$ (compare the case $k = 2$ in Section~\ref{sec:geo-spatial-multi}).
Hence, $s(n)$, the total number of special rectangles will satisfy the following recursively defined inequality:
  \begin{align*}
    s(n) &\leqslant 2km^k(m+1)^{k-1} + 2^k s(m).
  \intertext{Re-writing, we have}
    s(n) - 2^k s\left(\frac{n}{2}\right) &\leqslant 2k \left(\frac{n}{2}\right)^k \left(\frac{n}{2} + 1\right)^{k-1},
  \intertext{and applying the binomial theorem, we obtain}
    s(n) - 2^k s\left(\frac{n}{2}\right) & \leqslant 2k \left(\frac{n}{2}\right)^k \sum_{i=0}^{k-1} \binom{k-1}{i}\left(\frac{n}{2}\right)^i;
  \intertext{finally, applying Proposition~\ref{pro:recurrence}, we have}
    s(n) &\leqslant 2k \left(\frac{n}{2}\right)^k \left(\log n + \sum_{i=1}^{k-1} \binom{k-1}{i}\frac{n^i-1}{2^i-1}\right).
  \end{align*}
Now the maximum out-degree of any special hyperrectangle is $2^k$, so \mbox{$e(n) \leqslant 2^k s(n)$}; and $\displaystyle\frac{n^i-1}{2^i-1} \leqslant \frac{n^{i+1}-1}{2^{i+1}-1}$ for all $n \geqslant 2$ and all $i \geqslant 1$.
Hence, we have
  \[
    e(n) \leqslant 2k n^k \left( \log n + \frac{n^{k-1}-1}{2^{k-1}-1} \sum_{i=1}^{k-1} \binom{k-1}{i}\right) = 2k n^k \left(\log n + n^{k-1}-1\right),
  \]
as required.
\end{proof}

There is little related work on interval-based access control for arbitrary dimensions.
\citeN{yuan:effi09}, whose work was briefly described in Section~\ref{sec:geo-spatial}, stated that their methods for geo-spatial access control could be generalized to higher dimensions, without providing any details.
\citeN{sriv:scal08} generalized the notion of binary encryption trees to geo-spatial access control and higher dimensions.
The number of keys and the number of key derivation steps required are $\bigO{2^{k+1}\log n}$ where $k$ is the number of dimensions.
The schemes we describe in this paper are the first in which each user has a single key.

%% file: conclusion.tex
\section{Concluding Remarks} \label{sec:conclusion}

In this paper we consider the enforcement of an interval-based access control policy, which generalizes the temporal and geo-spatial access control policies in the literature~\cite{atal:effi07,atal:inco07,desa:new08}.
Such policies can be enforced using cryptographic methods, often called key assignment schemes.
There are several efficient key assignment schemes in the literature, in the sense that the amount of storage and the time taken to derive cryptographic keys is considerably less than that required if standard enforcement schemes are applied directly.
These efficient schemes exist because of the particular structure of the graph that is used to represent interval-based access control policies.
Existing work has used generic techniques for reducing the diameter of the graph, without considering the particular relationship between the access control policy and the desired graph.

In this paper we have developed a number of efficient enforcement schemes that have considerable advantages over existing ones.
We focus on the development of novel techniques to provide efficient schemes designed specifically for interval-based access control policies, rather than using more generic techniques.
Our approach enables us to produce, in almost all cases, exact values for the number of edges and the number of steps required to derive a key, in contrast to existing work in the literature (as shown by Figure~\ref{tbl:comparison}).
Moreover, we demonstrate that our constructions can be generalized to higher dimensions, yielding new insights into the efficient cryptographic enforcement of interval-based access control policies.

One disadvantage of our work in Section~\ref{sec:interval} is that we assumed that each dimension contained intervals in $T_n$ for some fixed $n$.
In practical applications, this may not be a reasonable assumption, and it may be prohibitively expensive to ``pad'' each dimension and work with $T^k_N$, where $N = \max\set{n_1,\dots,n_k}$ and $n_i$ is the number of points in the $i$th dimension.
One important aspect of our future work, therefore, will be to try to extend our results in Section~\ref{sec:interval} for $T^k_n$ to the more general case $T_{n_1} \times \dots \times T_{n_k}$.

Perhaps the most interesting area for future work is to consider more expressive access control policies and their enforcement using cryptographic techniques.
At the moment, we consider intervals defined over a totally ordered set of attributes $A$.
We also intend to consider policies where $A$ is some partially ordered set and users and objects are associated with subsets of $A$.
Of particular interest is the case where $A$ is a powerset defined over some set of attributes, since the resulting policies would be analogous to those used in ciphertext-policy attribute-based encryption~\cite{beth:ciph07}.
We have recently published some preliminary results in this area~\cite{cram:fast10}.

\paragraph*{Acknowledgements}
The author would like to thank Alex Dent, George Loizou, Keith Martin and the anonymous referees for their helpful comments.

%% file: appendix.tex
\section{Extending our constructions}\label{app:extended-temporal}

We now consider what modifications are required to our work to enable the derivation of keys for all nodes in $T_m$ (rather than just leaf nodes).
We do this to demonstrate that we can extend our constructions to the cases that are considered by previous researchers.

We first consider the following problem.
Given a diamond-shaped grid $D_m$ (illustrated in Figure~\ref{fig:d-8-log-derivation}), how can we reduce the diameter of the graph by adding a small number of edges?
Clearly, the number of edges in the grid is $(m-1)^2$ and the diameter of the graph is $2m-2$ for the set of edges shown in Figure~\ref{fig:d-8-log-derivation}(a).
Consider the following construction.

\begin{Sch}
\quad
\begin{enumerate}
  \item Divide $D_{2m}$ into four copies of $D_m$.
  \item Label these copies $\dialabel{m}{0}{0}$, $\dialabel{m}{0}{1}$, $\dialabel{m}{1}{0}$ and $\dialabel{m}{1}{1}$ (bottom to top and left to right, as illustrated in Figure~\ref{fig:d-8-log-derivation}).
  \item For each node in $[x,y] \in \dialabel{m}{1}{1}$ we define a set of nodes $C_{x,y}$, such that
          \[
            \card{C_{x,y} \cap \dialabel{m}{i}{j}} =  1\quad \text{and}\quad \card{C_{x,y}} = 4.
          \]
        Specifically, for all $i$ and $j$, the largest node in $\dialabel{m}{i}{j}$ that is less than or equal to $[x,y]$ belongs to $C_{x,y}$.
  \item For each node $[x,y] \in \dialabel{m}{1}{1}$ we connect the four nodes in $C_{x,y}$ together using $4$ edges.
        In doing this, we can get from any node in $C_{x,y}$ to another element of $C_{x,y}$ in a different copy of $D_m$ in no more than two hops.
  \item We now apply this construction recursively to each copy of $D_m$.
\end{enumerate}
\end{Sch}

An enforcing edge set for $D_8$ of diameter $3$ is shown in Figure~\ref{fig:d-8-log-derivation}(b).
The dashed lines represent the edges connecting copies of $D_2$; each copy of $D_2$ is labeled in the manner described above.

\begin{figure}[h] \centering
  \subfigure[The basic edge set]{\begin{minipage}{.45\textwidth}\centering\includegraphics{d_8_grid.md}\end{minipage}}
    \hfill
  \subfigure[An enforcing edge set of diameter $3$]{\begin{minipage}{.45\textwidth}\centering\includegraphics{d_8_log_derivation.md}\end{minipage}}
\caption{$D_8$}\label{fig:d-8-log-derivation}
\end{figure}

From this construction we deduce the recurrence relation
  \begin{align*}
    E_D(m) &= 4\left(\frac{m}{2}\right)^2 + 4E_D\left(\frac{m}{2}\right) \\
           &= m^2 + 4E_D\left(\frac{m}{2}\right),
  \end{align*}
where $E_D$ represents the number of edges required for $D_m$.
And from this recurrence relation we can prove by induction that $E_D(m) = m^2\log m$.
Let $d(m)$ be the diameter of the graph $(D_m,E)$, where $E$ is the edge set obtained from the above construction.
Then we have $d(m) = 2 + d(m/4)$, from which we may deduce that $d(m) = \log m$.

We can now construct an enforcing set of edges for $T_m$.
Consider the following construction.
\begin{Sch}
\quad
\begin{enumerate}
  \item Divide $T_{2m}$ into two copies of $T_{m}$, labeled $\trlabel{m}{0}$ and $\trlabel{m}{1}$ and a diamond $D_m$.
  \item For each node $[x,y] \in D_m$ define $C_{x,y}$ to be the following set of nodes:
    \begin{itemize}
      \item $[x,y]$;
      \item the largest element in $\trlabel{m}{0}$ that is less than $[x,y]$; and
      \item the largest element in $\trlabel{m}{1}$ that is less than $[x,y]$.
    \end{itemize}
  \item For each node $[x,y] \in D_m$ add two edges to connect it to the other two nodes in $C_{x,y}$.
        Hence, it is now possible to get from any node in $D_m$ to a node in each copy of $T_m$ in one hop.
  \item Apply this construction recursively to $\trlabel{m}{0}$ and $\trlabel{m}{1}$.
  \item Construct a set of edges for each of $\dialabel{m}{0}{0}$, $\dialabel{m}{0}{1}$, $\dialabel{m}{1}{0}$ and $\dialabel{m}{1}{1}$.
\end{enumerate}
\end{Sch}

From this construction we deduce the recurrence relation
  \begin{align*}
    E_T(m) &= 2\frac{m^2}{4} + E_D\left(\frac{m}{2}\right) + 2E_T\left(\frac{m}{2}\right) \\
           &= \frac{1}{2}m^2 + \frac{1}{4}m^2(\log m - 1) + 2E_T\left(\frac{m}{2}\right) \\
           &= \frac{1}{4}m^2(\log m + 1) + 2E_T\left(\frac{m}{2}\right),
  \end{align*}
where $E_T(m)$ denotes the number of edges required for $T_m$.

\begin{Pro}
For all $n \geqslant 2$, $E_T(n) = \frac{1}{2}n^2\log n$.
\end{Pro}

\begin{proof}
By inspection, we have $E_T(2) = 2$, so the result holds for $n=2$.
Now assume that the result holds for all $n < N$.
Then
  \begin{align*}
    E_T(N) &= \frac{1}{4}N^2(\log N + 1) + 2E_T\left(\frac{N}{2}\right) \\
          &= \frac{1}{4}N^2(\log N + 1) + 2\left(\frac{N^2}{8}(\log N - 1)\right) \quad \text{by inductive hypothesis} \\
          &= \frac{1}{2}N^2\log N
  \end{align*}
as required.
\end{proof}

\section{Multi-key constructions for temporal access control}\label{app:multi-key-temporal}

\subsection{The case $k = 3$}

\begin{figure}[h] \centering
  \includegraphics[width=.95\textwidth]{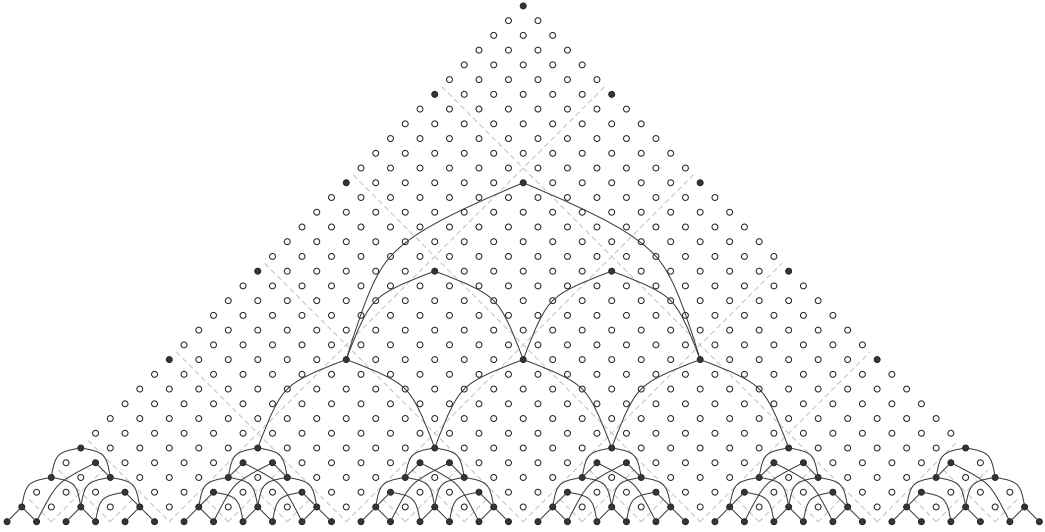}
\caption{The key derivation graph generated by applying Construction~\ref{sch:3-key-construction} to $T_{36}$}\label{fig:3-key-scheme-t36}
\end{figure}

Suppose that $m = ab$.
Then we can treat $T_m$ as a triangle $T_b$ in which the leaf supernodes are copies of $T_a$ and the non-leaf supernodes are copies of $D_a$.
Now any interval $[x,y]$ can be represented as the union of no more than three intervals $[x,z_1]$, $[z_1+1, z_2]$,  $[z_2+1,y]$, where $z_1$ and $z_2$ are multiples of $a$.
Then $[z_1+1,z_2]$ has the form $[z'_1 a + 1, z'_2 a]$, for some integers $z'_1$ and $z'_2$, and can be represented as the interval $[z'_1,z'_2]$ in $T_b$.
Figure~\ref{fig:3-key-scheme-t36} illustrates $T_{36}$ split into copies of $T_6$ and $D_6$.
The interval $[3,25]$, for example, can be represented as $[3,6] \cup [7,24] \cup [24,25]$, and $[7,24]$ can be treated as the interval $[1,3]$ in a copy of $T_4$ comprising interior copies of $D_6$.

Hence, we only require key derivation edges between the ``interior'' maximal nodes in each diamond supernode and the nodes in each triangle supernode.
An exterior maximal node is of the form $[1,z'_1 a]$ or $[z'_2 a + 1, m]$: $[1,z'_1 a]$ is the union of the intervals $[1,a]$ and $[a+1,z'_1 a]$; and $[z'_2 a + 1, m]$ is the union of the intervals $[z'_2 a + 1, (b-1)a]$ and $[(b-1)a+1,m]$.
In other words, all intervals corresponding to maximal nodes in an exterior diamond supernode can be treated as the union of a leaf supernode and an interior diamond supernode.
%
%
Clearly we can apply this construction recursively, yielding the following construction.

\begin{Sch}\label{sch:3-key-construction}
Let $m = a_1\dots a_k$.
  \begin{enumerate}
    \item Treat $T_m$ as a tree $T_{a_1}$ comprising supernodes $T_{m/a_1}$ and $D_{m/a_1}$.
    \item Mark every node in the upper edges of each copy of $T_{m/a_1}$ as a special node (as in Construction~\ref{sch:2-key-construction}).

          More formally, denote the $i$th leaf supernode by $T^{(i)}_{m/a_1}$, $1 \leqslant i \leqslant a_1$.
          Then
            \[
              T^{(i)}_{m/a_1} = \set{[x+(i-1)m/a_1,y+(i-1)m/a_1] : 1 \leqslant x \leqslant y \leqslant m/a_1},
            \]
          and we define the set of special nodes to be
            \[
              \bigcup_{i=1}^{a_1-1} \set{[x,(i-1)m/a_1,im/a_1] : 1 \leqslant x \leqslant m/a_1} \cup \bigcup_{i=1}^{a_1-1} \set{[im/a_1+1,y] : 1 \leqslant y \leqslant m/a_1}
            \]
    \item Mark the maximal node in each interior copy of $D_{m/a_1}$ as a special node.
          This set of nodes has the form $\set{[im/a_1+1,jm/a_1] : 1 < i \leqslant j < a_1}$ and hence these nodes are in one-to-one correspondence with $T_{a_1-2}$ (under the mapping $[im/a_1+1,jm/a_1] \mapsto [i-1,j-1]$).
          Construct a set of key derivation edges for this set of nodes by applying binary decomposition to $T_{a_1 - 2}$.
    \item If $m/a_1 \geqslant 2$, repeat for all leaf supernodes $T_{m/a_1}$.
  \end{enumerate}
Finally, define every key allocation edge that connects two special nodes to be a key derivation edge.
\end{Sch}

The application of this construction to $T_{36}$ is illustrated in Figure~\ref{fig:3-key-scheme-t36}.
The faint dashed lines illustrate the recursive partitioning of $T_{36}$ and copies of $T_6$ into supernodes.

Now Step 1 adds no more than $2m$ special nodes\footnote{In fact, Step 1 adds precisely $m/a_1 + (a_1 - 2)(2m/a_1 - 1) + m/a_1$ special nodes.} and Step 2 adds $\frac{1}{2}(a_1-1)(a_1-2)$ special nodes.
Hence, the number of special nodes satisfies the inequality
  \[
    s(m) \leqslant 2m + \frac{1}{2}(a_1-1)(a_1-2) + a_1 s\left(\frac{m}{a_1}\right)
  \]
Now for $a_1 \leqslant \sqrt{m}$ and $a_{i+1} \leqslant \sqrt{a_i}$ we have $s(m) \leqslant \frac{5}{2}m + a_1 s(m/a_1)$ and we can easily prove by induction that $s(m) \leqslant \frac{5}{2}m\log\log m$.
By construction, the out-degree of each node in the key derivation graph is $2$, so $e(m) \leqslant 5m\log\log m$.

Moreover, the number of hops $d(m)$ satisfies the inequality
  \[
    d(m) \leqslant  \ceil{\log (\sqrt{m} - 2)} + d(\sqrt{m}) < \frac{1}{2}\log m + 1 + d(\sqrt{m}).
  \]
Hence, we may conclude that $d(m) \leqslant \log m + \ceil{\log\log m}$.


\subsection{The case $k \geqslant 4$}
For $k = 3$, we ensure that we can derive keys for the maximal nodes in each diamond supernode using a single key.
When $k \geqslant 4$, we simply require that we can derive keys for the maximal nodes using $k-2$ keys.


Let $s(m,k)$ denote the number of special nodes required to construct a scheme with $k$ keys for $T_m$.
Then we have
  \[
    s(m,k) \leqslant 2m + s(a_1 - 2,k-2) + a_1s(m/a_1,k)
  \]
Consider $k = 4$: let $a_1 = m/\log m$ and $a_{i+1} = a_i/\log a_i$, and recall that the number of special nodes required by a $2$-key scheme for $T_m$ is no greater than $m\log m$.
Then we have
  \begin{align*}
    s(m,4) &\leqslant 2m + \frac{m}{\log m}\log\left(\frac{m}{\log m}\right) + \frac{m}{\log m}s(\log m,4) \\
           &< 3m + \frac{m}{\log m}s(\log m,4).
  \end{align*}
From this inequality, we prove by induction that $s(m,4) \leqslant 3m\log^* m$.
Clearly the result holds for $m = 4$.
Suppose, then, that $s(m,4) \leqslant 3m\log^* m$ for all $m < N$.
  \begin{align*}
    s(N,4) &\leqslant 2N + s\left(\frac{N}{\log N},2\right) + \frac{N}{\log N} s(\log N, 4) \\
           &\leqslant 2N + \frac{N}{\log N}\log \left(\frac{N}{\log N}\right) + \frac{N}{\log N}\underset{\text{by inductive hypothesis}}{\underbrace{3 \log N \log^*(\log N)}} \\
           &< 3N + 3N \log( \log^* N) \\
           &= 3N(1 + \log (\log^* N)) \\
           &= 3N \log^* N
  \end{align*}
Hence, we require no more than $6m\log^* m$ edges to construct a $4$-key scheme for $T_m$.
We also have
  \[
    d(m) \leqslant \ceil{\log (m/\log m)} + d(\log m) \leqslant \log m - \log\log m + 1 + d(\log m),
  \]
and there are at most $\ceil{\log^* m}$ recursive steps, so (applying a similar argument to the one used for $k=3$) we have $d(m) \leqslant \log m + \ceil{\log^* m}$.
We summarize the results of this section in the following theorem, which is stated without proof.

\begin{Thm}
There exist enforcing sets of edges $E_1$, $E_2$ and $E_3$ such that
  \begin{itemize}
    \item $\card{E_1} = 2m\log m$, the graph $(T_m,E_1)$ comprises two disconnected components, and the diameter of $(T_m,E_1)$ is $\floor{\log m}$;
    \item $\card{E_2} = 5m\log\log m$, the graph $(T_m,E_2)$ comprises three disconnected components, and the diameter of $(T_m,E_2)$ is $\log m + \ceil{\log\log m}$;
    \item $\card{E_3} = 6m\log^* m$, the graph $(T_m,E_3)$ comprises four disconnected components, and the diameter of $(T_m,E_1)$ is $\log m + \ceil{\log^* m}$;
  \end{itemize}
\end{Thm}